\documentclass[12pt,preprint]{aastex}
\usepackage{emulateapj5} 
\usepackage{psfig} 

\def\ltsima{$\; \buildrel < \over \sim \;$} 
\def\simlt{\lower.5ex\hbox{\ltsima}} 
\def\gtsima{$\; \buildrel > \over \sim \;$} 
\def\simgt{\lower.5ex\hbox{\gtsima}} 
  
\received{04/13/2001} 
\accepted{} 
\journalid{}{} 
\articleid{}{} 
\slugcomment{The Astrophysical Journal, 567:1-18, 2002 March 1} 
\lefthead{} 
\righthead{} 
 
\begin{document} 
    
\title{Temperature Profiles of Nearby Clusters of Galaxies} 
\author{Sabrina De Grandi} 
 \affil{Osservatorio Astronomico di Brera, Via Bianchi 46, 
I-23807 Merate (LC), Italy} 
\email{degrandi@merate.mi.astro.it} 
 
\and 
 
\author{Silvano Molendi} 
 \affil{Istituto di Fisica Cosmica, CNR, via Bassini 15, 
I-20133 Milano, Italy} 
\email{silvano@ifctr.mi.cnr.it} 
 
\begin{abstract} 

We report results from the analysis of 21 nearby galaxy clusters, 11
with cooling flow (CF) and 10 without cooling flow, observed with
BeppoSAX.  The temperature profiles of both CF and non-CF systems are
characterized by an isothermal core extending out to $\sim
0.2~r_{180}$; beyond this radius both CF and non-CF cluster profiles
rapidly decline.  Our results differ from those derived by other
authors who either found continuously declining profiles or
substantially flat profiles.  Neither the CF nor the non-CF profiles
can be modeled by a polytropic temperature profile, the reason being
that the radius at which the profiles break is much larger than the
core radius characterizing the gas density profiles.  For $r >
0.2~r_{180}$, where the gas can be treated as a polytrope, the
polytropic indices derived for CF and non-CF systems are respectively
$1.20\pm0.06$ and $1.46\pm0.06$.  The former index is closer to the
isothermal value, 1, and the latter to the adiabatic value, 5/3.
Published hydrodynamic simulations do not reproduce the peculiar shape
of the observed temperature profile, probably suggesting that a
fundamental ingredient is missing.

\end{abstract} 
 
 
\section {Introduction} 
 
Galaxy clusters, being the largest virialized systems in the universe, 
are useful cosmological probes. Through the assumption of hydrostatic 
equilibrium for the X-ray emitting intra-cluster medium (ICM) it is 
possible to measure the total gravitational mass of clusters 
(including the dominating dark matter component) as a function of the 
gas temperature and density profiles. 
 
Current models of structure formation suggest that the mass components
of clusters are representative of the universe as a whole.  Once the
gas mass is determined from X-ray data in the deprojection or fitting
analysis (see e.g. Fabian et al. 1981; Ettori \& Fabian 1999) and the
total mass is estimated through the hydrostatic equilibrium, it is
possible to derive the gas fraction.  This fraction is a lower limit
on the global baryon fraction and can be used to constrain the
cosmological density parameter if combined with primordial
nucleosynthesis calculations (e.g. White et al. 1993).  If well
calibrated, the slope and evolution of cluster scaling relations, such
as the gas mass vs. temperature (e.g. Finoguenov, Reiprich \&
B\"ohringer 2001; Voit 2000) and the cluster size vs. temperature
(e.g. Mohr et al. 2000; Verde et al. 2001), can be used to constrain
cosmological and structure formation models . For instance, the
mass-temperature relation is fundamental in linking theoretical
cosmological models, which give the mass function of clusters via the
Press-Schechter formalism, and the observed temperature function
(e.g. Henry 1997; Nevalainen, Markevitch \& Forman 2000). 
Temperature profiles are also fundamental in determining the gas 
entropy distribution (see Lloyd-Davies, Ponman \& Cannon 2000 and 
references therein), which is a powerful tool to explore 
non-gravitational processes that could alter the specific thermal 
energy in the ICM, such as pre-heating of the gas before it falls into 
the cluster core and energy injection from supernova-driven galaxy 
winds. 
 
While ROSAT X-ray images allowed high-precision studies of the gas 
density distribution in individual clusters (e.g. Mohr, Mathiensen \& 
Evrard 1999), spatially resolved temperature measurements have 
only become possible recently with ASCA and BeppoSAX. 
With previous missions (e.g. HEAO1, {\it Einstein}, EXOSAT, Ginga)
only global temperature measurements were available and therefore the
ICM was assumed to be roughly isothermal. ROSAT was not the ideal
instrument to measure temperature structures in rich clusters given its
soft spectral range (0.1-2.4 keV), nonetheless various attempt have
been made in the last years for a few very bright clusters (e.g. Briel
\& Henry 1994, 1996; Knopp, Henry \& Briel 1996; Davis \& White 1998)
or using color ratios techniques (Irwin, Bregman \& Evrard 1999;
Sanders, Fabian \& Allen 2000).
 
Different studies have found conflicting results regarding  
temperature gradients in clusters.  In 1998 Markevitch and 
collaborators (hereafter MFSV98) analyzed azimuthally averaged radial 
temperature profiles for 30 clusters observed with ASCA, finding that 
nearly all clusters show a significant radial temperature decline at 
large radii. By fitting the composite temperature profile for 
symmetric clusters with a polytropic relation up to 
$\sim 55\%$ of the virial radius, MFSV98 found that the 
temperature decline corresponded to a polytropic index of $\sim 1.24$ on 
average.  However, doubts on the universality and the steepness of 
this profile have been raised by subsequent studies. 
Irwin et al. (1999) comparing the MFSV98 results with the work from 
other authors (see references in Irwin et al. 1999), found isothermal 
temperature profiles, even for those clusters where MFSV98 found a 
temperature decline. Moreover, from an investigation of ROSAT PSPC 
color profiles, they found that clusters were generally isothermal, 
even in clusters in common with the MFSV98 analysis. 
After the analysis of a sample of 106 clusters observed with 
ASCA, White (2000) concluded that $90\%$ of the temperature profiles 
were consistent with being flat. 
For a sample of cool clusters (i.e. kT$\lesssim 4$ keV) Finoguenov et 
al. (2001) derived ASCA/SIS temperature profiles similar to the 
universal temperature profile of MFSV98, although the radial range 
explored by these authors was smaller than that studied by MFSV98. 
 
Although ASCA has been the first X-ray instrument able to perform
spatially resolved spectroscopy in hot clusters given its adequate
energy range (1-10 keV), its large and strongly energy-dependent
point-spread function (PSF) required complicated correction procedures
for the spectral analysis of extended sources.  Different works on
temperature measurements with ASCA have applied different methods to
correct for the PSF effects.  In this respect BeppoSAX (Boella et
al. 1997a) is a more suitable instrument to investigate temperature
structures in galaxy clusters and supplies an independent dataset with
respect to ASCA. The Medium-Energy Concentrator Spectrometer (MECS;
Boella et al. 1997b) on board BeppoSAX works in a similar energy
range of ASCA, but has a sharper PSF (HPR $\sim 1^\prime$), which is
radially symmetric and almost energy independent (D'Acri, De Grandi \&
Molendi 1998).
 
Recent results for several nearby clusters (e.g. A2319: Molendi et
al. 1999; A3266: De Grandi \& Molendi 1999a; A2256: Molendi, De Grandi
\& Fusco-Femiano 2000; A3562: Ettori et al. 2000; A3571: Nevalainen et
al.  2001; etc.)  show that BeppoSAX has the ability to obtain better
constraints than ASCA on temperature profiles.  Irwin \& Bregman
(2000) (hereafter IB00), who analyzed a sample of 11 clusters observed
with BeppoSAX, placed further arguments supporting a general
isothermality of the ICM. IB00 claimed that the temperature profiles
were generally flat or increase slightly out to $\sim 30\%$ of of the
virial radius.
 
In this paper we investigate radial cluster temperature profiles for a
sample which doubles the IB00 sample using new archival and
proprietary BeppoSAX data. Contrary to IB00, who limited the analysis
within $9^\prime$ (i.e. to regions $\lesssim 30\%$ of the virial
radius), we will consider the full field of view of the MECS in order
to extend our radial temperature analysis out to $\sim 50\%$ or more
of the virial radius (i.e.  to radii comparable to those explored by
MFSV98 and White 2000).  Using an independent dataset from ASCA we are
interested in searching if temperature declines are a common
phenomena in clusters, and if this is the case, what is the typical
shape or gradient. We will also test if declines are specific to
certain types of systems.
 
The plan of this paper is as follows. In Section 2 we describe the
sample.  In Section 3 we present the BeppoSAX observations and discuss
in details the X-ray data analysis, the spectral modeling is presented
in $\S$ 3.1 and a technical comparison with the IB00 sample is in $\S$
3.2. In Section 4 we investigate the projected temperature profiles
derived for our sample, in $\S$ 4.1 we test the temperature profiles
at large radii for instrumental or systematic effects, in $\S$ 4.2 and
4.3 we apply various models to the profiles and discuss some
implication of the results.  In Section 5 we compute the averaged
cluster temperature profiles and compare them to previous measures
($\S$ 5.1), to XMM-Newton observations ($\S$ 5.2) and to
hydrodynamical cluster simulations ($\S$ 5.3). A summary of our main
conclusions is given in Section 6.
 
Quoted confidence intervals are $68\%$ for 1 interesting parameter 
(i.e. $\Delta \chi^2 =1$), unless otherwise stated. 
We use $H_o=50$ Mpc km s$^{-1}$ and $q_o=0.5$. 
 
\section {The Sample} 
 
The aim of this paper is to derive spatially resolved temperature 
measurements for a sample of rich and nearby (i.e., $0.02\lesssim z 
\lesssim 0.1$) clusters of galaxies observed with the BeppoSAX 
satellite and to investigate the possible 
existence of an universal temperature profile for rich clusters.   
We selected all the clusters with on-axis pointings and exposure times 
larger than 30 ks which were available at the BeppoSAX SDC archive at 
the end of March 2001 (including a few proprietary observations). 
 
We have divided the sample of the 21 selected clusters into a
subsample of 10 non-cooling flow (non-CF) and 11 cooling flow (CF)
objects on the basis of the ROSAT analysis presented in Peres et
al. (1998), defining a non-CF cluster an object with a mass deposition
rate consistent with zero.
 
Our redshift range allows us to explore in details the innermost regions 
of a few nearby clusters (i.e. Coma, A1367, Perseus and A3627) out to 
$\sim 20\%$ of the virial radius, and to measure temperatures out to 
$\sim 50\%$ of the virial radius for the other clusters in the sample. 
 
A detailed discussion of the metal abundance profiles derived for a
subset of 17 clusters of this sample is presented in De Grandi \&
Molendi (2001). In a forthcoming paper (Ettori et al. 2001 in prep.)
we will discuss the mass profiles derived for this sample. The
observation log for the current cluster sample is given in Table 1.
 
\section {Data Analysis} 
 
In this paper we have considered observations from the MECS on board
BeppoSAX. The MECS is presently composed by two units working in the
1-10 keV energy range, with energy resolution $\sim 8\%$ and angular
resolution $\sim 0.7^{\prime}$ (FWHM), both computed at 6 keV.
 
For each cluster we have analyzed the data from the MECS2 and MECS3 
separately.  Standard reduction procedures and screening criteria have 
been applied using the SAXDAS package under FTOOLS environment to 
produce equalized and linearized MECS event files.  Using the 
information contained in the housekeeping files we have rejected all 
events which have occurred at times when the instantaneous 
pointing direction differed by more than $10^{\prime\prime}$ from the 
mean pointing direction. 
 
The PSF of the MECS instrument is known to vary only weakly with
energy (D'Acri et al. 1998).  Although we expect PSF-induced spectral
distortions to be small, we have properly taken them into account
creating corrected effective area files with the EFFAREA program
available within the SAXDAS package. This program convolves the ROSAT
PSPC surface brightness profile, computed from archival PSPC pointings
data, with an analytic model of the MECS PSF to estimate spectral
distortions.  EFFAREA also includes corrections for the
energy-dependent telescope vignetting for on-axis observations.
 
Each cluster has been divided into concentric annuli centered on the 
X-ray emission peak. Out to $8^{\prime}$ we accumulated spectra from 4 
annular regions each $2^{\prime}$ wide, beyond this radius we 
accumulated spectra from annuli $4^{\prime}$ wide.   
Whenever it is necessary we exclude pointlike sources along the line  
of sight by excising a circular region with radius $2^\prime$.   
For each cluster the radial profile stops at the last annulus 
containing more than $30\%$ source counts with respect to the total 
(i.e.  source$+$background) counts. 
The dominant contribution to the MECS background at energies larger
than $\sim$ 5 keV is from events induced by the interaction of high
energy particles with the structure surrounding the instrument. Using
data acquired during occultations of the satellite from the dark earth,
Vecchi et al. (1999) have monitored the non X-ray background finding
that variations are typically contained within $\sim 5\%$ from the
mean.  In the present work, we have decided to account for these
variations by excluding from our analysis spectra from the outermost
regions not satisfying the conditions on the intensity of the source
with respect to the background indicated above.
We believe that this is preferable to the alternative choice of using
this data and including a systematic component to the error budget to
account for possible variations in the background, the reason being
that if such a component exists and is truly systematic it will show
up again when we average our profiles.
For the spectra which satisfy the conditions on the intensity of the
source with respect to the background indicated above, fluctuations of the 
background of up to $\sim 5\%$ do not introduce a significant
increase in the error of the temperature measurement.

The energy range used for the spectral fitting is 2-10 keV with the 
exceptions described in the following. 
In the outermost annuli if the source counts drops to less than 
$50\%$ with respect to the total counts, we restrict the energy range 
to 2-8 keV, to avoid possible distortions from the hard MECS 
instrumental background (see the above discussion). 
 
One of the most important steps in the MECS data reduction is the
correct treatment of the detector entrance window structure.  The
Beryllium window is sustained by a thicker supporting structure, the
so called strongback\footnote{see {\it
http://sax.ifctr.mi.cnr.it/Sax/Mecs/tour.html}}, 
in form of a circular
ring and four ribs.  The transmission of the Be window is function
both of the energy (being the prime responsible of the low energy cutoff
in the effective area), and position (because of the presence of the
strongback).  The strongback affects the detector regions covered by
its geometrical shadow convolved through the detector PSF and produces
an artificial hardening in the spectra accumulated from these regions.
Taking into account this convolution the circular ring of the
strongback starts at $\sim 8^\prime$ and ends at $\sim 12^\prime$ from
the detector center.  We have computed the corrected effective area
for the $8^\prime-12^\prime$ annulus by considering the typical
thickness of the strongback and its transmission as a function of the
energy and position. Moreover, in the $8^\prime-12^\prime$ annulus we
restrict our analysis to the range 3.5-10 keV to avoid the low energy
part of the spectrum where our correction is less reliable.  All other
regions of the detector covered by the strongback have been
appropriately masked and the data rejected. 
Another problem related to the strongback is that the detector center 
and the pointing axis of the X-ray telescope are not coincident, and 
that the distance between these two points, albeit small, is 
different in MECS2 and MECS3. 
Thus the same region of an extended source (in sky coordinates) is
shadowed by the strongback differently in the two MECS. Therefore, it
is important to apply the strongback correction to the effective areas
of the two MECS separately.
 
The background subtraction has been performed using spectra extracted 
from blank sky events files in the same region of the detector as the 
source. 
Total and background spectra and effective areas of the two MECS units 
have been added together using the FTOOLS {\it mathpha} and {\it 
addarf}, respectively. 
 
\subsection{Spectral modeling} 
 
We have rebinned (FTOOLS {\it grppha}) the spectra to collect a 
minimum of 25 counts in each energy bin to allow the use of the 
$\chi^2$ analysis.  All spectral fits have been performed using XSPEC 
version 10.00. 
 
We fit each observed spectra with a thermal emission model for a 
low-density plasma in collisional ionization equilibrium ({\it mekal} 
model in XSPEC), absorbed by the nominal Galactic column density 
(Dickey \& Lockman 1990; {\it wabs} model). This model assumes thermal 
emission at a single temperature and is characterized by three free 
parameters, the gas temperature, the metal abundance given relative to 
the solar value, and the normalization.  We find that freeing the 
redshift to account for the systematic shift of $\sim 40$ eV in the 
MECS conversion from channel to energy affect the temperature 
measurements by less than $5\%$ (in agreement with the analysis of 
IB00). 
 
Recent observations from Chandra (Hydra A: David et al. 2000)
and XMM (M87, A1795 and A1835: Molendi \&
Pizzolato 2001) X-ray observatories show a lack of spectroscopic
evidence for multiphase gas in the core of cooling flow clusters.
Therefore we will not apply the multiphase models that we (e.g. De
Grandi \& Molendi 1999b; Molendi \& De Grandi 1999) and others
(e.g. Allen et al. 2001 and references therein) have used in the past,
and will use single temperature models for all regions.
 
\subsection {Comparison with another BeppoSAX sample} 
 
We compare here the results on radial temperature profiles derived by
Irwin \& Bregman (2000) (IB00) from another sample of galaxy clusters
observed with BeppoSAX.  This sample consists of 11 clusters found in
the BeppoSAX archive, 9 of which are in common with our sample.  We
will concentrate in this Section on the technical differences between
our X-ray data analysis and that performed by IB00. A comparison
between the scientific results is presented in Section 5.1.
 
The analysis performed by IB00 differs from ours in several 
respects. The most important difference is that these authors limit 
the radial analysis at a radius of $9^\prime$, whereas our analysis 
considers the whole useful detector area up to $20^\prime$. The second 
difference is that IB00 do not correct the data for the strongback 
effect, claiming that this effect becomes a factor only at radii 
larger than $9^\prime$.  

This is not the case because the strongback starts effecting the spectra 
at about 8$^\prime$ from the center of the detector and also because the 
center of the cluster can be offset by up to 2$^\prime$ from the 
center of the detector (for further details see Section 3). 
Finally, they do not investigate the difference between CF and non-CF 
clusters in such a systematic way as we do. 
 
IB00 find that the temperature profiles of their clusters are
generally flat, or increase slightly out to $\sim 30\%$ of the virial
radius and that a decline of the temperature of $14\%$ out to $30\%$
of the virial radius is ruled out at the $99\%$ confidence level.
These results appear to be in contradiction with the ones we report in
Section 4. To further investigate this difference we have considered
three CF clusters in common between our sample and that of IB00
(i.e. A85, A496 and 2A0335$+$096), and we have performed the analysis
of the X-ray data in the same way as described in IB00. Namely, we
have used the 3.5-10.5 keV energy band and we have not applied any
correction for the strongback.  As shown in Figure 1, where we compare
our results against those of IB00, we are able to reproduce the
temperature profiles derived by IB00. We note that the profiles from
both analysis are slightly increasing with the radius.
  
In Figure 2 we compare the temperature profiles for the same three
clusters derived using our procedure, which includes a treatment for
the strongback (see Section 3 for details), to those of IB00.  We find
our profiles to be in broad agreement with those of IB00.  In all
three clusters the temperature measured in the outermost bin by IB00
is slightly larger than the mean temperature of our two overlapping
bins.  This difference can be most likely attributed to the lack of a
proper treatment of the strongback by IB00.  Moreover, by extending
our analysis further out wee see that the profiles start to decline.
 
\section {Projected Radial Temperature Profiles}  
 
In Figure 3 and 4 we present the radial temperature profiles for the 
non-CF and CF clusters, respectively, plotted against the radius in 
units of $r_{180}$, which is the radius encompassing a spherical 
density contrast of 180 with respect to the critical density. In an
$\Omega=1$ cosmology this radius approximates the cluster virial radius.
We compute $r_{180}$ from the prediction given by the simulations of 
Evrard, Metzler \& Navarro (1996): $r_{180}$ = 3.95 Mpc ($T_{X}/10$ 
keV)$^{1/2}$. Here $T_{X}$ is the mean emission-weighted temperature, 
in units of keV, which we have estimated by fitting the temperature 
profile with a constant up to the maximum radius available for 
each cluster. 
For the CF clusters we have computed the mean emission-weighted 
temperature from the temperature profile excluding the cooling flow 
region (i.e. a central circular region with radius equal or larger than 
the cooling radius).  From the X-ray spatial analysis of Peres et 
al. (1998) the derived cooling radii of the CF clusters in our sample 
are generally smaller than 2$^\prime$.  Exceptions are A2199 and 
2A0335$+$096, with cooling radii larger than 2$^\prime$ but smaller 
than 4$^\prime$, and the Perseus cluster with a cooling radius of 
$\sim 6^\prime$. For these clusters we have excluded the appropriate 
cooling flow regions. 
 
In Table 2 we report for each cluster the temperature measurement from
each annular region, the mean temperature $T_{X}$ with the
corresponding best-fit $\chi^2$ and degrees of freedom, the cluster
redshift and derived value for $r_{180}$.

\subsection {Testing Temperature Profiles at Large Radii} 

From a visual inspection of Figure 3 and 4 it is evident that a 
temperature gradient is present at large radii. 
We have performed a series of checks to test whether this temperature 
decline is driven by an instrumental or systematic effect. 
 
\subsubsection {Criterion on the Outermost Bin} 

We have applied a more conservative criterion to define the
outermost bin for the temperature profiles by imposing that the source
counts must exceed $40\%$ (instead of $30\%$ as described in Section
3) of the total counts.  By applying this restriction we find that in
the case of non-CF clusters the only bin excluded is the last
temperature bin of A3266. In the case of CF clusters the outermost
bins of 6 clusters (i.e. A1795, A2142, A2199, A3562, 2A0335$+$096 and
PKS0745$-$191) are excluded.  We have verified that none of the
results reported in the next Sections are substantially modified if we
reduce the sample by excising the above bins.

\subsubsection {Differences between Distant and Nearby Clusters} 

We have plotted in Figure 5 temperature profiles against the 
radius in arcminutes for four clusters observed at different 
redshifts.  Two of them are nearby cluster, i.e. Coma at z=0.02223 and 
A3627 at z=0.01628, which are observed by BeppoSAX out to radius of 
0.74 Mpc and 0.552 Mpc, respectively. The other two clusters, A119 at 
z=0.04420 and A2256 at z=0.0570, are more distant and we are able to 
determine a temperature out to 1.14 Mpc and 1.44 Mpc, respectively. 
We find that the temperature profiles of the two distant clusters 
decline with increasing radius, while those of the nearby systems 
remain constant. This implies that the temperature decline we observe 
in Figures 3 and 4 is not due to a systematic effect such as, for 
example, an incorrect vignetting correction at large off-axis angles. 
 
\subsubsection {Independent Temperature Measurements for A3562}

For one cluster belonging to our sample (A3562) we have two
independent BeppoSAX observations, the first one centered on the
cluster emission peak and the second one centered on the galaxy group
SC 1329$-$313, which is located at about $27^\prime$ west of the
emission peak of A3562.  We find that the two independent measurements
of a specific region of A3562 obtained from different parts of the
detector give consistent results. More specifically, Ettori et
al. (2000) derived the temperature map of A3562 finding a temperature
of $3.9^{+1.1}_{-0.8}$ keV in the annulus $8^\prime-16^\prime$ in the
sector pointing towards SC 1329$-$313 (sector west in Figure 9 of
Ettori et al. 2000).  This temperature is in agreement with the one
measured (Bardelli et al. 2001 submitted) from the observation
centered on SC 1329$-$313 in the same region (i.e.  $3.5\pm 0.3$ keV
in the $12^\prime-20^\prime$ annulus of the north-east sector reported
in Bardelli et al. 2001).

\subsubsection {Background Subtraction Effects} 

The increasing importance of background relative to source counts
could lead to systematic errors in the measured temperature in the
outermost bins. We have performed the following exercise to test this
possibility. We have taken the data for the two best-observed clusters
that do not show any evidence for a decrease out to $20^\prime$
(i.e. Perseus and Coma), and randomly extracted 3.5 ks of data for
Perseus and 8 ks for Coma.
Similarly, from the blank-sky event files we have randomly extracted
100 ks of data, which we have then added to the 3.5(8) ks of
Perseus(Coma) data.
In this way we have obtained a simulated ``faint'' Perseus(Coma)
cluster observed for about 100-110 ks.
The above exposure times have been chosen so that the number of source
counts and the number of source counts with respect to the total
counts in the $8^\prime-12^\prime$, $12^\prime-16^\prime$, and
$16^\prime-20^\prime$ bins is about the same as in more distant
clusters such as A2142, A1795 and A3562.
We have repeated this exercise three times for both clusters, each
time choosing different chunks of source and background data.

The six simulated data sets have been analyzed like the original data
as detailed in Section 3, the resulting temperature profiles for the
``faint'' Perseus and Coma clusters are shown in Figure 6.
We find that these profiles do not show any significant temperature
decrement up to the last useful bin indicating that the background
subtraction and/or low count rates are not leading to the temperature
decline seen in the other clusters of our sample.

\subsubsection {PSF and Vignetting correction effects} 

We have analyzed the clusters in our sample using the corrected and
uncorrected on-axis effective areas to check the entity of the
PSF and vignetting corrections at all radii and all temperature ranges. 
In the analysis with the uncorrected effective areas we have not
performed any strongback correction in the $8^\prime-12^\prime$ bin.  

In Figure 7 we report the case of Coma, Perseus and A1367 clusters.
Let us start by considering the measurements at small radii
($r<8^\prime$). In the case of A1367, which has a low temperature and
a relatively flat surface brightness profile neither vignetting nor
PSF corrections are important. For Perseus which has a somewhat larger
temperature and a rapidly declining surface brightness profile the PSF
correction slightly increases the temperature for bins from $2^\prime$
to $8^\prime$. For Coma, which has the highest temperature and almost
flat surface brightness profile, where the dominating effect is that
of the vignetting, the analysis with the corrected effective areas
leads to a slightly larger temperature.

For the $8^\prime-12^\prime$ bins we always find a higher temperature
in the uncorrected effective areas case because of the artificial
hardening of the spectrum introduced by the strongback (details are
given in Section 3). 

At large radii ($r>12^\prime$) for low temperature objects such as
A1367 no substantial effects are found, for higher temperature objects
(Perseus and Coma) the vignetting effect becomes progressively more
important, in the case of Perseus ($\sim 7$ keV) the vignetting
corrected temperature is $\sim 15-20\%$ higher than the uncorrected
one, while for Coma ($\sim 9$ keV) the vignetting corrected
temperature is $\sim 20-23\%$ higher than the uncorrected one.

In conclusion, for more then half of the points reported in Figure 7
the correction is smaller than $10\%$, only in the
$8^\prime-12^\prime$ bin and for the outermost bins of hot clusters
(kT$>6-7$ keV) the corrections span between $10\%-25\%$.

\subsubsection {The Break Radius}

As detailed in Section 4.2 our mean temperature profile is well
represented by a broken line model (see equation 1). In the following
we test whether the break could result from some systematic effect.
To this end we have computed, for each cluster, the break radius in
arcminutes, $r^{\prime}_b$, by fitting the temperature profile (note
that the slope of the line was fixed to the mean value, i.e. -2.56 and
-1.13, for the non-CF and CF clusters respectively, see Section 4.2 and
Table 3).
We have then compared $r^{\prime}_b$ to $r^{\prime}_{180}$,
i.e. $r_{180}$ expressed in arcminutes (see Figure 8). If the break
in the profile results from a systematic effect it is likely to occur
roughly at the same place in the detector for all clusters, on the
contrary if it reflects a genuine property of our objects it will
correlate with $r^{\prime}_{180}$. From Figure 8 we see that
$r^{\prime}_b$ is not the same for all clusters, and that clusters
with larger $r^{\prime}_{180}$ tend to have larger
$r^{\prime}_b$. Indeed, by fitting the data in Figure 8 first with a
constant and than with a line we find that the latter fit provides a
substantial improvement with respect to the prior (significance $>
99.5\%$ according to the F-test).
 
\subsection {Modeling of the Temperature Profiles} 

We have fitted all the temperature measurements reported in Figure 3
(non-CF clusters), with a constant finding, as expected, that this
simple model does not provide a good description of the data (see
Table 3).
If we try to reproduce the decline observed in Figure 3 by modeling 
the data with a line we find that the fit improves significantly ($> 
99.9\%$ according to the F-test). Nonetheless the large $\chi^2$ (see 
Table 3) implies that the fit is still very poor, most likely because 
the temperature decline is not continuous but begins at a radius of 
about $0.2~r_{180}$. We have therefore modeled our data with a 
broken line defined as: 
$$ t = t_b \quad {\rm for}~ x < x_b \quad $$ 
$$ t = t_b + m*(x - x_b) \quad {\rm for}~ x > x_b \quad  \eqno(1)$$ 
where $t \equiv T/T_X$ and  $x \equiv r/r_{180}$ are respectively the 
normalized temperature and radius. The break radius $x_b$, 
the temperature in the isothermal region $t_b$, and the slope of 
the line $m$ for $x > x_b$ are the free parameters of the 
model.  This model provides a better description of the data ($> 
99.9\%$ significance according to the F-test) when compared to the 
simple line model.  

We have performed the same set of fits for the temperature
measurements of the CF clusters plotted in Figure 4, results are
reported in Table 3.  As for the non-CF clusters a line
provides a better fit than a constant (significance $> 99.9\%$
according to the F-test) and a broken line provides a better fit than
a line (significance $> 99.9\%$ according to the F-test).  
Interestingly, while the best fitting $90\%$ confidence for the break
radius $x_b$ for the two samples overlap, the best fitting slopes $m$
appear to be different at more than the $99.99\%$ confidence level
({\bf $5.5\sigma$}). 
Since the break radius and the slope parameter are correlated we have
further investigated the difference in $m$ by performing a fit to the
non-CF and CF clusters with the value of the break radius fixed at
0.20 (the mean break radius obtained when simultaneously fitting CF
and non-CF clusters). The slope for the non-CF clusters is found to be
larger at the $99.95\%$ significance level ({\bf $3.7\sigma$}) than the
one for CF clusters, implying that the temperature profile of the
former decreases more rapidly than that of the latter. This issue will
be discussed in greater detail in Section 4.3.

Given the similarity of the non-CF and CF cluster profiles we have 
joint the two samples and performed fits to the total sample.  Results 
are reported in Table 3, not surprisingly they are similar to 
those derived for the two individual subsamples.   

\subsubsection {The Break Radius}

In the following we investigate the properties of the break radii of
the individual temperature profiles of our clusters.
For each cluster we have computed the break radius by fitting the
broken line model described in equation (1), to the temperature
profile.
We have performed two sets of fits, in the first the slope of the
broken line model was fixed to the mean value (i.e. -2.56 for non-CF
and -1.13 for CF clusters), while in the second it was left as a free
parameter.
In the following we use the values of the break radius derived by
fixing the slope.  By allowing the slope to be a free parameter we
find break radii with larger uncertainties with respect to those
derived by fixing the slope although the values are consistent each
others within $\sim 1\sigma$. This is because the analysis of the
individual temperature profiles is often limited by the poor
statistics.
Note that in the case of Coma, A3627 and A2319, where the profiles are
flat, the fitting algorithm pushes the value of the break radius
beyond the last data-point. In this case we have fixed the value of
the break radius to the radius of the last data-point and have
determined the lower bound by varying the break radius until the best
fitting $\chi^2$ was increased by 1 with respect to the value found by
fitting the temperature profile with a constant; the upper bound of
the break radius is of course unconstrained.  For all profiles with
the exception of A754, which is an extremely disturbed and
substructured cluster (e.g. Blinton et a. 1998; Henriksen \&
Markevitch 1996; Zabludoff \& Zaritsky 1995), the broken line model
provides an acceptable fit to the data, implying that virtually all
our profiles can be adequately represented by a broken line model.

The large-scale distribution of the gas temperature in Perseus
cluster has been studied with multi-pointing ASCA observations by
Ezawa et al. (2001) and Dupke \& Bregman (2001).
The latter authors derived temperatures from circular regions of $20^\prime$
radius at about $40^\prime$ from the Perseus cluster emission
peak, finding that in the outer regions the gas is roughly isothermal
with temperatures of $\sim 6-7$ keV.
The profile found by Ezawa et al. (2001) is in agreement with ours out
to $20^\prime$. In the $20^\prime-28^\prime$ annular region the
authors find a temperature of $\sim 8$ keV, beyond this region their
temperature profile shows a smoothly decline of $\sim 2$ keV out to
$60^\prime$.  This decline is similar to what we expect for Perseus
on the basis of our broken-line model
(i.e. a decline of about 2.5 keV from $20^\prime$ to $60^\prime$),
however the normalization found by Ezawa et al. (2001) is higher
by about 1-2 keV.

In Figure 9 we show the break radius in units of $r_{180}$ as a
function of redshift for all the objects in our sample.  In some cases
(e.g. A1367 and 2A0335$+$096) the break radius is well constrained, while
in others (e.g. PKS0745$-$191 and A2199) our data does not allow a firm
measurement.  The mean value of $x_b$ is 0.20 with a standard
deviation of 0.068.
We have performed a statistical test to verify if the above dispersion
is due completely to the measurements uncertainties or if it is in part
due to an intrinsic dispersion in the distribution of break radii.
From the test (Maccacaro et al. 1988), which assumes that the
parent population of break radii is distributed like a Gaussian with a
mean value, $\overline x_b$, and an intrinsic dispersion,
$\sigma_{x_b}$, we find that the intrinsic dispersion is not
consistent with zero, 
$\sigma_{x_b} = 3.1^{+1.4}_{-0.9}\times 10^{-2}$ (errors are
at the $68\%$ confidence limit for one interesting parameter).  
We conclude that, while the broken line model can adequately fit all
our temperature profiles, the break radius is not the same for all
objects, although the intrinsic dispersion is relatively small.

Interestingly, if we divide our sample in CF and non-CF cluster and
redo the analysis described above, we find that while for CF clusters
the intrinsic dispersion is consistent with 0, for non-CF clusters it
is not at more than the $99\%$ significance level.  Such a difference
could result from the fact that CF systems are closer to virialization
and therefore more similar one to the other while non-CF systems are
irregular and therefore more different one from the other.

\subsection {The Polytropic model for the mean Temperature Profile}

The broken line model is only a phenomenological model with no
physical basis. We have therefore tried comparing our data with a
polytropic model.  As discussed in Markevitch et al. (1999) and Ettori
(2000), under the assumptions that the gas density profile is well
represented by a $\beta$-model and that a polytropic relation holds
between the gas density and the real three dimensional temperature,
the projected temperature profile can be expressed as: 
$$ t = t_o ( 1 + x^2/{x_c}^2)^{-1.5\beta(\gamma - 1)} \eqno(2)$$
   
where $t \equiv T/T_X$, $x \equiv r/r_{180}$, $ x_c \equiv
r_c/r_{180}$, $\beta$ is the well known $\beta$ parameter, $\gamma$ is
the polytropic index and $t_o$ is the normalized temperature at $x=0$.
We have assumed a gas distribution with $\beta = 2/3$.
In the following we describe the application of this model to the mean
temperature profile over the full radial range covered by our data.

\subsubsection {The full radial range}

We performed a first fit leaving $t_o$, $x_c$ and $\gamma$ as free
parameters in equation (2), with the last parameter constrained
between the two limiting values of 1 (isothermal gas) and 5/3
(adiabatic gas).  The best fit (see Table 4) is found for $\gamma =
5/3$ and for $x_c = 0.48^{+0.04}_{-0.02}$ which, when converted into
$r_c$ by using the $r_{180}$ value obtained from the mean temperature
of the clusters reported in Table 2, yields $r_c =
1.49^{+0.06}_{-0.13}$ Mpc. This value is unacceptable as it is many
times larger than the core radius derived by fitting the surface
brightness profiles of galaxy clusters.  We have attempted a second
fit fixing $x_c = 0.08$, which corresponds to $r_c = 0.25$ Mpc.  The
best fitting value of $\gamma$ is $1.09\pm0.01$, implying an almost
isothermal profile, the $\chi^2$ value is large implying that the fit
is very poor, indeed the flat model profile does not reproduce the
drop observed at radii larger than about 0.2 $r_{180}$.  We conclude
that the temperature profile for our non-CF clusters is inconsistent
with a polytropic temperature model.
 
The fits with a polytropic law of the CF clusters yield results
similar to those found for the non-CF clusters.  From the first fit,
where $x_c$ is a free parameter, we derive a value of $r_c$ which is
many times larger than the core radius obtained by fitting surface
brightness profiles of galaxy clusters.  From the second fit we find a
best fitting value of $\gamma$ of $1.06\pm0.01$, implying an almost
isothermal profile which, as for the non-CF clusters, does not
reproduce the decline observed at radii larger than about 0.2.  We
note that the $\chi^2$ for the second fit is in itself not
unacceptable, however, if we compare this fit with the broken line
fit, or with the polytropic fit with free $x_c$, which are both
characterized by a substantial temperature decline at large radii, we
find that they provide a significantly better description of the data
(in both cases at more than the $99.99\%$ confidence level according
to the F-test).
 
We have also performed fits to the total sample, comprising both
non-CF and CF clusters, results are reported in Table 4.  
In summary our main results are that the polytropic model fails to fit
the data adequately and that the broken line model is the one which
gives the best description of the data.

\subsubsection{The outer regions}

Outside about $0.2~r_{180}$ both CF and non-CF clusters are 
characterized by a clear temperature decline.  A power-law profile 
gives acceptable fits for the temperature measurements of CF and 
non-CF clusters for $r > 0.2~r_{180}$.  The results of the fits, 
which are reported in Table 5, show that the CF clusters have a 
smaller power-law index than the non-CF systems, implying that the 
temperature gradient is shallower in the first than in the latter. 
 
Under the assumption that the three-dimensional temperature profile can be
adequately described by a power-law of the form ${T(r) = T_o
r}^{-\mu}$, the parameters describing the power-law (i.e. the slope
$\mu$ and the normalization $T_o$ ) can be derived analytically
from the parameters of the power-law fitting the observed projected
temperature profile, indeed the two power-law slopes are the same (see
Appendix A for details).  Let us also assume that the gas density
profile for $r > 0.2~r_{180}$ can be described by a power-law of
the form $n(r) = n_o r^{-\nu}$, with slope $\nu =
2$, as is expected if the gas density profile follows a $\beta$-model
with $\beta=2/3$.  Then the polytropic index, defined from
the relationship $ p \propto n^{\gamma}$, where $p$ is the gas pressure,
is simply given by $ \mu /2 + 1$.  
Taking $\mu$ from the best fits reported in
Table 5 we find that $\gamma = 1.20 \pm 0.06 $ for CF systems and
$\gamma = 1.46 \pm 0.06 $ for non-CF systems. We recall that $\gamma =
1$ corresponds to the isothermal case and $\gamma = 5/3$ describes the
adiabatic case.  Both values are contained within the two limiting
cases, with the CF systems being closer to the isothermal value and
non-CF systems to the adiabatic value. 

A possible interpretation is that since for non-CF systems the time
since the last major merger is smaller than for the CF systems, heat
transport processes will have had more time to act on the CF systems
than on the non-CF systems and their temperature profiles will be
flatter.  
It may well be that after a merger the ICM is left in a convectively
unstable state.  If this is the case, the gas will assume an adiabatic
configuration in a relatively short time, $\lesssim 10^9$ Gyr (the
time scale on which convection operates is of the order of the sonic 
time scale). Once the gas has become adiabatic, conduction will set in
as the dominant heat transport mechanism. In the outer regions
conduction operates on time scales comparable to the Hubble time (see
discussion in Section 5.4.2 of Sarazin 1988). This would explain why,
albeit reduced, the temperature gradient persists in the ``older'' CF
systems.
 
\section {The Mean Temperature Profiles}  
 
In Figure 10 we plot the mean error-weighted temperature profiles 
for the CF and non-CF clusters. 
The two profiles appear to be quite similar. There are however two 
differences: the temperature for the CF clusters in the innermost bin 
is much lower that the one for non-CF clusters, indicating that cooler gas 
is present in the core of CF clusters; the decline observed 
at radii larger than $\sim 0.2~r_{180}$ in both subsamples appears 
to be more rapid in non-CF systems than in CF systems, a possible 
interpretation for this difference is discussed in Section 4.3. 
  
\subsection {Comparison with other Cluster Samples} 
 
In this subsection we compare our profiles with similar profiles
derived by other authors.
In Figure 10 we report the result found by IB00 for the cluster sample
already introduced in Section 3.2.  IB00 found that their normalized
temperature profiles are flat out to $0.2~r_{180}$, and that from
0.2 to 0.3 $r_{180}$, the profiles rise somewhat. The model that
provides the best-fit to the data is the linear model which is
overplotted in Figure 10 with its $90\%$ confidence range of the
slope. Moreover, IB00 found that a temperature drop of $14\%$ from the
center out to $30\%$ of the virial radius can be ruled out at the
$99\%$ confidence level.  By inspecting Figure 10 we can firstly
note that the IB00 analysis failed to observe the temperature decline
which is already starting at about $20\%$ of the virial radius. The
reason for this is most likely the inadequate treatment of the
strongback effects in the IB00 analysis of the BeppoSAX data, as
detailed in Section 3.2.
Our profiles show that the temperature at $0.3~r_{180}$ is smaller than
the temperature in the second bin by about $20\%$ (we do not consider
the first bin as it is affected by the cooling flow). An increase of
$7\%$, similar to the one found by IB00, is excluded on the basis of
our profiles at more than the $99.99\%$ significance level.

MFSV98 from the analysis of ASCA data, found that almost all clusters
in their sample show a temperature decrease with radius and that, when
plotted in the same normalized units as in Figure 10, they are
remarkably similar.  In particular they find that for a 7 keV cluster
with a typical gas density profile, the observed drop in temperature
can be characterized by a polytropic index of $1.2-1.3$.
In Figure 10 we overlay on our results the composite temperature
profile derived for symmetric clusters by MFSV98. The long-dashed box
encloses the scatter of their best-fit values, whereas the dotted one
encloses approximately all their temperature profiles and most of the
associated error bars (see Figure 8 in MFSV98).
When comparing the boxes and our mean profiles it is clear that the
large uncertainties in the ASCA composite profile lead to a rough
agreement between the two results.
However, the characteristic shape of the temperature profiles found by
our BeppoSAX analysis is missed in the MFSV98 result.
While the outermost regions of both composite profiles are in fair
agreement showing a similar temperature decline, in the innermost
regions of the clusters they are different as in the MFSV98
profile there is no evidence of the isothermal core.

In another work based on ASCA observations of a sample of 106
clusters, which includes all those in the MFSV98, White (2000)
concluded that temperature profiles are generally consistent with
isothermality.
The significance of this result is such that when cluster's
temperature profiles are fitted by power-law functions, approximately
$90\%$ are consistent with isothermality at the $3\sigma$ limit (see
Figure 5 in White 2000).
Interestingly, for those clusters in common with MFSV98, White found
that often his core temperatures are cooler and his outer temperatures
are hotter than MFSV98. 
However the core temperatures found by White (2000) are derived from
single-temperature fits, whereas those found by MFSV98 are ambient
temperatures for the cooling flow fits.
Although the radial range explored by White (2000) is similar to ours
and to that of MFSV98, his large systematic uncertainties on
temperature profiles prevent to really constrain these profiles at
large radii.

\subsection{Comparison with XMM Observations}

For the two clusters in our sample where temperature profiles derived
from XMM-Newton data are published, i.e. Coma cluster (Arnaud et
al. 2001a) and A1795 (Arnaud et al. 2001b, Tamura et al. 2001), we have
overplotted the XMM-Newton profile on our BeppoSAX profile (after
converting their $90\%$ uncertainties to our $1\sigma$ confidence
limits). In the case of A1795 (see Figure 11 upper panel), if we
exclude the cooling-flow region, where the superior angular resolution
of XMM-Newton allows a much more detailed measurement of the
temperature profile, the BeppoSAX and XMM-Newton profiles are in
remarkably good agreement. We note that while the XMM-Newton profile
goes out to 12$^{\prime}$ the BeppoSAX profile extends to
16$^{\prime}$, this difference results from the better sensitivity of
BeppoSAX to the low surface brightness emission arising from the outer
region of this cluster. Although the effective area of XMM-Newton 3
units adds up to roughly 2200 cm$^2$ at 2 keV and 1300 cm$^2$ at 6 keV
while the two operating MECS units on BeppoSAX total 70 cm$^2$ at 2
keV and 75 cm$^2$ at 6 keV, the background, which is dominated by the
instrumental component, is characterized by an intensity per unit
solid angle 30 to 50 times higher in one MOS unit than in one MECS
unit.  Consequently the sensitivity of XMM-Newton to low surface
brightness sources in the 2-10 keV band is actually inferior to that
of BeppoSAX.

In the case of Coma (see Figure 11 lower panel), if we exclude the
innermost arc-minute, where the XMM-Newton profile drops because of
softer emission coming from the galaxy NGC 4874, the BeppoSAX and
XMM-Newton profiles run parallel. Both profiles are slightly declining
with radius and the systematic difference is of the order of 0.8 keV.
The most likely cause for such a difference is a modest error in the
effective areas at high energies of either BeppoSAX or
XMM-Newton. Without going into unnecessary details it will suffice to
say that while the effective area calibration of BeppoSAX is
considered to be quite solid and stable, the high energy effective
area calibration of XMM-Newton is still underway.
     
\subsection{Comparison with Hydrodynamic Cluster Simulations} 
 
In this Section we compare our averaged temperature profile to radial
temperature profiles derived from hydrodynamic cluster simulations.
The mean observed profile shown in Figure 12 (circles) has been
computed by including all temperature measurements, except in the
innermost bin, where we do not include CF cluster measurements.

The projected emission-weighted temperature profiles derived from
hydrodynamic simulations plotted in Figure 12, have been all computed
from the three-dimensional profiles following eq. (A2) and (A3) in the
Appendix under the hypotheses that the $\alpha$ parameter in eq. (A2)
is equal to 0.4 (Ettori 2000), and that the gas density profile is
described by a $\beta$-model, $ n(r) = n_o
[1+(r/r_c)^2]^{-{3\over2}\beta}$, with $r_c = 0.25$ Mpc and $\beta =
2/3$.
  
In Figure 12 we plot projected temperature profiles computed by
Evrard, Metzler \& Navarro (1996).  These simulations have been
designed to explore the effects of galactic winds on the structures of
the ICM in a standard Cold Dark Matter (CDM) $\Omega=1$
cosmogony. Here we report three different cases: an $\Omega =1$
without winds, an $\Omega =1$ universe with winds, and an open
$\Omega=0.2$ universe without winds.  The three-dimensional profiles
have been normalized by the authors using the X-ray emission-weighted
temperature.  As shown in Figure 12 the main characteristic of all
these profiles is that in the outer cluster regions they are flatter
than the observed profile.  These profiles show a pronounced
flattening towards the cluster center at a radius similar to that
where we observe a flattening in the observed profile, but this effect
is probably due to the limited resolution of the simulation.
  
A simulation with a better resolution was performed by Eke, Navarro \& 
Frenk (1998) for a flat, low-density, $\Lambda$-dominated CDM universe 
($\Omega=0.3$, $\Lambda=0.7$).
In Figure 12 we plot the projected gas temperature profile computed 
from the three-dimensional radial profile averaged over 10 cluster 
simulation at $z=0$ (Eke et al. 1998). The authors have appropriately 
normalized this profile scaling the individual temperature profiles to 
the virial temperature of each system. 
When comparing in Figure 12 their temperature profile with the observed 
one, we find that the simulated profile is too steep for radii smaller 
than $\sim 0.2~r_{180}$ and that it is too flat for radii larger 
than $\sim 0.2~r_{180}$, failing to reproduce the observed profile 
at all radii. 
 
Bialek, Evrard \& Mohr (2001) have systematically investigated the
effects of ICM pre-heating on the basic observed properties of the
local cluster population. Bialek et al. have kindly provided us with a
projected temperature profile to be compared with our mean temperature
profile (see Figure 12).  Their profile is the average of twelve
cluster simulations performed for a $\Lambda$-CDM cosmology
($\Omega=0.3$, $\Lambda=0.7$) characterized by an initial entropy
level of 105.9 keV cm$^{-2}$ (entropy level S3 in Bialek et al. 2001),
and it is normalized to match our observed core temperature.  From
Figure 12 we find again that this profile is too flat at large radii
to reproduce the observed profile.

The last profile that we have considered is obtained from the 
Santa Barbara cluster comparison project (Frenk et al. 1999), which 
simulated the formation of a single X-ray cluster in an 
$\Omega=1$ universe using twelve nearly-independent codes. 
As widely discussed in Frenk et al. (1999) the codes show agreement in 
reproducing the various cluster properties ($\sim 10\%$ agreement for 
the ICM properties), therefore we will consider in the following the 
averaged mass-weighted gas temperature profile only (solid line in 
Figure 17 in Frenk et al. 1999), which was obtained by the authors 
averaging the data calculated by each code.   
To normalize the temperature profile we have used the mean temperature 
computed by fitting with a constant model the profile between $0 - 
1.3$ Mpc, which is the radial range covered on average by the clusters 
observed in our sample. Using this temperature we have also computed 
$r_{180}$ as given in Section 4 to normalize the cluster radius. 
From Figure 12 we firstly note a difference between the Evrard et 
al. (1996) profile for an $\Omega=1$ universe and the Frenk et 
al. (1999) profile, which is computed in the same cosmological 
framework. The Frenk et al. (1999) profile rises rapidly towards the 
cluster center whereas the Evrard et al. (1996) profile is 
flatter. This difference is probably due to the smaller resolution of 
the code used by Evrard et al. (1996), as discussed in the Santa 
Barbara comparison paper. 
The difference between our observed temperature profile and that 
derived by Frenk et al. (1999) is substantial in the inner parts of 
the clusters where the simulated temperature profile is still rising 
at the innermost point plotted. 
In the outer parts of the cluster, both the observed and simulated 
temperature profile drop in fair agreement. 
 
In general, we can conclude that none of the simulations considered
here is able to reproduce the peculiar shape of the observed
temperature profile, i.e. the isothermal core in the inner cluster
regions followed by a steeply declining temperature profile towards
the outer regions.  This could suggest that a fundamental ingredient
is missing in the construction of the hydrodynamical simulations.
This ingredient could be a heat transport mechanism that rapidly
brings the ICM within a radius of $\sim 0.2~r_{180}$ to the same
temperature, indeed the conduction time scale is of the order of a few
Gyrs for $r < 0.2~r_{180}$ (see discussion in Section 5.4.2 of Sarazin
1988).
Alternatively, it might be that the merger itself acts as a heat
transportation mechanism.  Future N-body and theoretical works should
be able to reproduce this characteristic feature.
 
\section {Summary} 
 
We have performed spatially resolved temperature measurements for a sample 
of 21 rich and nearby galaxy clusters observed by BeppoSAX. 
Our sample comprises 10 non-CF and 11 CF clusters. 
Below we report our main findings. 
 
$\bullet$ 
The temperature profiles of both CF and non-CF systems are characterized 
by an isothermal core extending out to $\sim 0.2~r_{180}$; beyond this 
radius both CF and non-CF cluster profiles declines. The temperature drops 
by a factor of almost 2 from $r \sim 0.2~r_{180}$ to $r\sim 0.5~r_{180}$. 
 
$\bullet$ 
Neither the CF nor the non-CF profiles can be modeled by a polytropic 
temperature profile, the reason being that the radius at which the 
profiles break is much larger than the core radius characterizing the 
gas density profiles. 
 
$\bullet$ 
For $r > 0.2~r_{180}$ both CF and non-CF temperature profiles can be
modeled by a power law, the CF systems have a flatter slope that the
non-CF systems. The polytropic indices derived from the power law
slopes are respectively $1.46\pm0.06$ for non-CF systems and
$1.20\pm0.06$ for CF systems.
Both values are contained within the isothermal ($\gamma=1$) and the 
adiabatic ($\gamma=5/3$) case, with the CF systems being closer to the 
isothermal value and the non-CF systems to the adiabatic value. 
A possible interpretation is that since for non-CF systems the time
from the last major merger is smaller than for the CF systems, heat
transport processes will have had more time to act on the CF systems
than on the non-CF systems and their temperature profiles will be
flatter.
 
$\bullet$ 
None of the previously published mean temperature profiles show the
characteristic shape, i.e. an isothermal core followed by a rapid
decline, that we find. The mean profile found by MFSV98, obtained from
ASCA data, shows a smooth decline with no evidence for an isothermal
core.  The profile found by IB00 from the analysis of 11 BeppoSAX
clusters IB00 features a rise of about $10\%$ when going from the
center to a radius of $0.3~r_{180}$.  This profile, as can be seen in
Figure 10, is in disagreement with ours.  As detailed in Section 3.2,
the reason for the difference is most likely the inadequate treatment
of the strongback effects in the IB00 analysis of the BeppoSAX data.

$\bullet$ 
None of the hydrodynamic simulations we have considered reproduces 
the peculiar shape of the observed temperature profile, i.e. the 
isothermal core in the inner cluster regions followed by the steep
temperature decline in the outer regions. This suggests that a fundamental 
ingredient is missing in the construction of the hydrodynamical simulations. 
Future N-body and other theoretical works should be able to reproduce 
this characteristic feature.

\acknowledgments 
 
R. Fusco-Femiano is thanked for allowing us to use proprietary data
(one observation of A754 and A119) prior to publication. The authors
thank S. Borgani, S. Ettori, S. Ghizzardi, F. Governato and
F. Pizzolato for useful discussions.  We acknowledge support from the
BeppoSAX Science Data Center.  Part of the software used in this work
is based on the NASA/HEASARC FTOOLS and XANADU packages.
 
  

\clearpage 

\begin{figure}
\epsscale{0.9}
\centerline{\psfig{file=f1.eps,width=3.5in,angle=-90}} 
\figcaption{Radial temperature profiles for 3 clusters contained both in
our sample and in that of Irwin \& Bregman (2000) (i.e. A85, A496 and
2A0335$+$096), as a function of the radius. Filled circles are
temperature measurements from Irwin \& Bregman (2000), open triangles
are derived by us analyzing the data in the same way as in Irwin \&
Bregman (2000) (see text for details).  All error bars are at
$1\sigma$ ($68\%$ confidence level); we have converted the $90\%$
errors reported in Figure 2 of Irwin \& Bregman (2000) into $68\%$
errors by dividing them by 1.65.  }
\vskip 1.cm
\epsscale{0.9}
\centerline{\psfig{file=f2.eps,width=3.5in,angle=-90}} 
\figcaption{Radial temperature profiles for 3 clusters contained both in
our sample and in that of Irwin \& Bregman (2000) (i.e. A85, A496 and
2A0335$+$096), as a function of the radius.  Filled circles are
temperature measurements from Irwin \& Bregman (2000), open circles
are temperatures derived from our analysis as described in Section 3.
}
\end{figure}

\clearpage 

\begin{figure}
\epsscale{0.8}
\centerline{\psfig{file=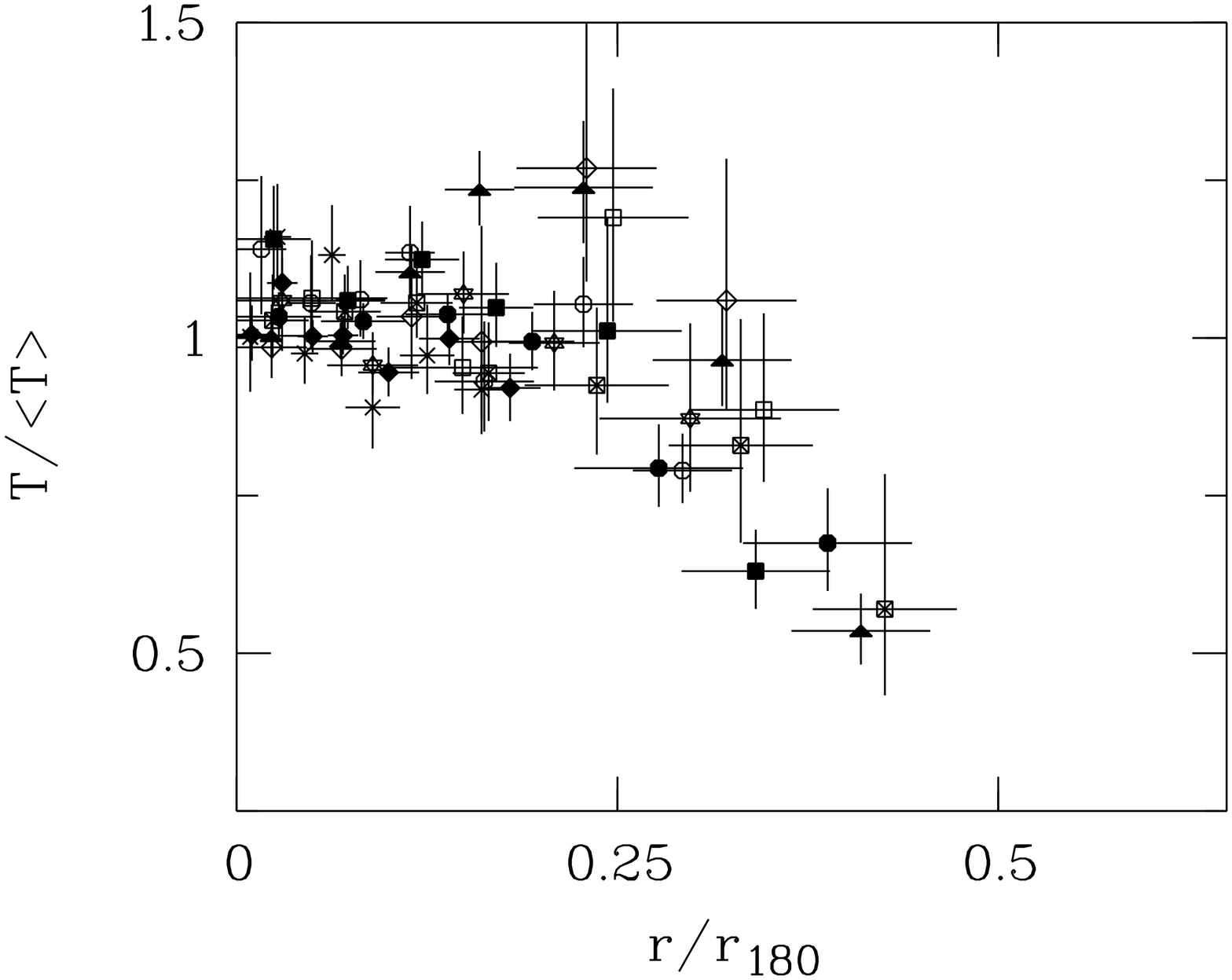,width=5.5in,angle=-90}} 
\figcaption {Temperature profiles (projected) for the non-CF clusters,  
plotted against radii in units of $r_{180}$.  Clusters are related to 
symbols as follows: A119 (filled squares), A754 (filled triangles), 
A1367 (open circles), A1750 (open squares), A2256 (filled circles), 
A2319 (open lozenges), A3266 (crossed squares), A3376 (stars), A3627 
(crosses) and Coma (filled lozenges). 
}
\vskip 1.cm
\epsscale{0.8}
\centerline{\psfig{file=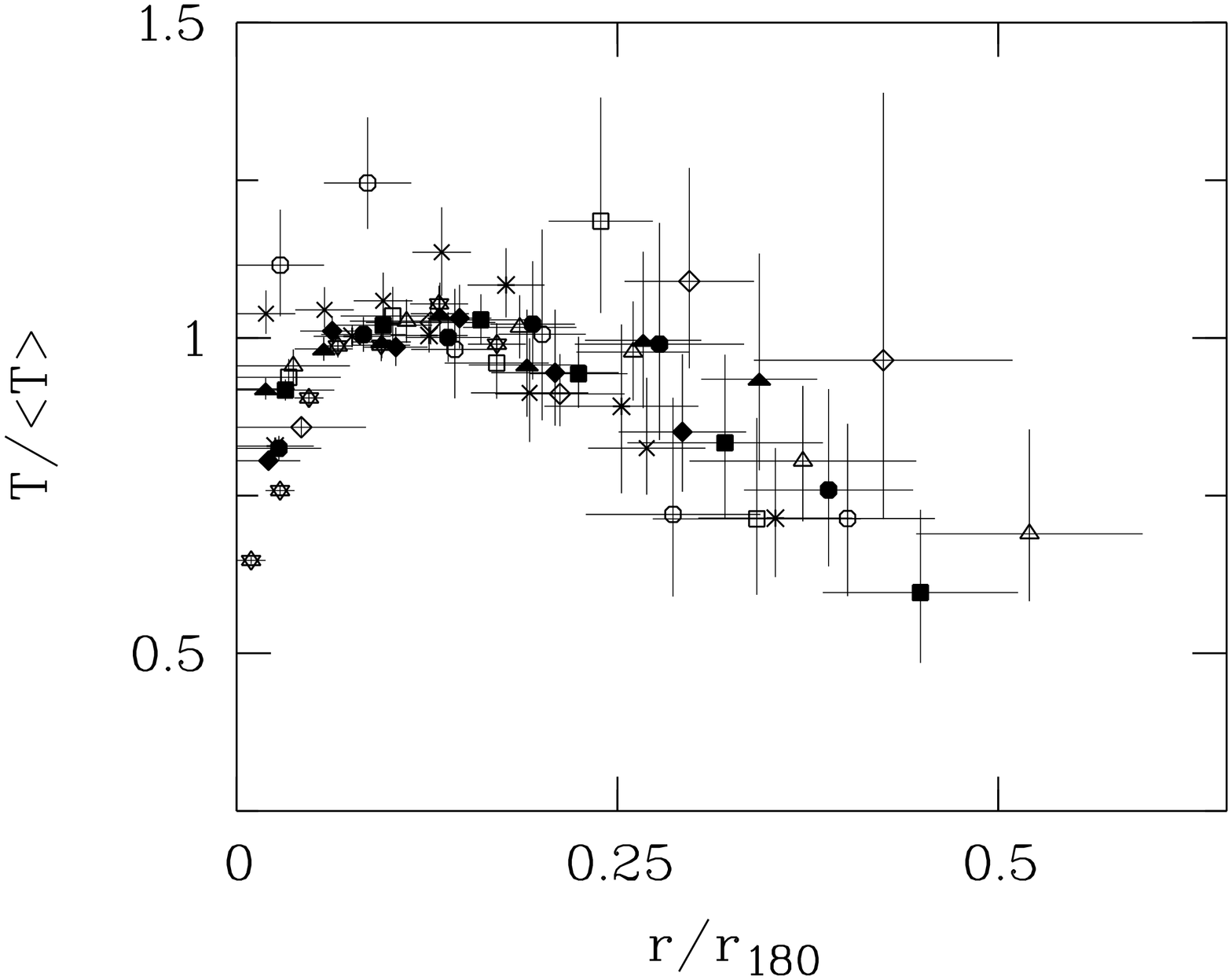,width=5.5in,angle=-90}} 
\figcaption{Temperature profiles (projected) for the CF clusters, plotted  
against radii in units of $r_{180}$. Clusters are related to symbols 
as follows: A85 (filled circles), A496 (filled lozenges), Perseus 
(stars), A1795 (filled squares), A2029 (open squares), A2142 (open 
triangles), A2199 (filled triangles), A3562 (open circles), A3571 
(crosses), 2A 0335$+$096 (asterisks) and PKS 0745$-$191 (open 
lozenges). }
\end{figure}

\clearpage 

\begin{figure}
\epsscale{0.8}
\centerline{\psfig{file=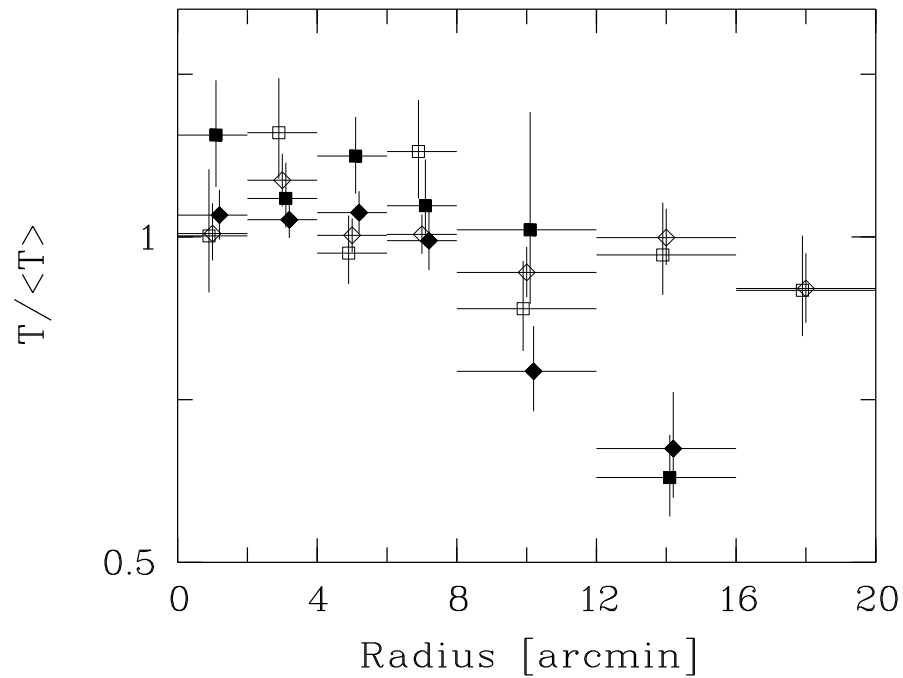,width=5.5in,angle=-90}} 
\figcaption {Temperature profile vs. radius in arcminutes for Coma  
(open lozenges), A3627 (open squares), A119 (filled squares) and A2256 
(filled lozenges). Coma and A3627 are nearby clusters at redshift about 
0.02, whereas A119 and A2256 are more distant clusters at redshift 
0.0442 and 0.0570, respectively.  
}
\vskip 1.cm
\epsscale{0.8}
\centerline{\psfig{file=f6.eps,width=5.in,angle=-90}} 
\figcaption{Temperature profiles vs. radius in arcminutes for 
``faint'' simulated Coma (a) and Perseus (b) clusters (details are
given in Section 4.1.4). Filled circles are the original BeppoSAX
temperature profiles for Coma and Perseus, open symbols are the
profiles resulting from six simulated data sets.}
\end{figure}

\clearpage 
 
\begin{figure}
\vskip -1.cm
\epsscale{0.8}
\centerline{\psfig{file=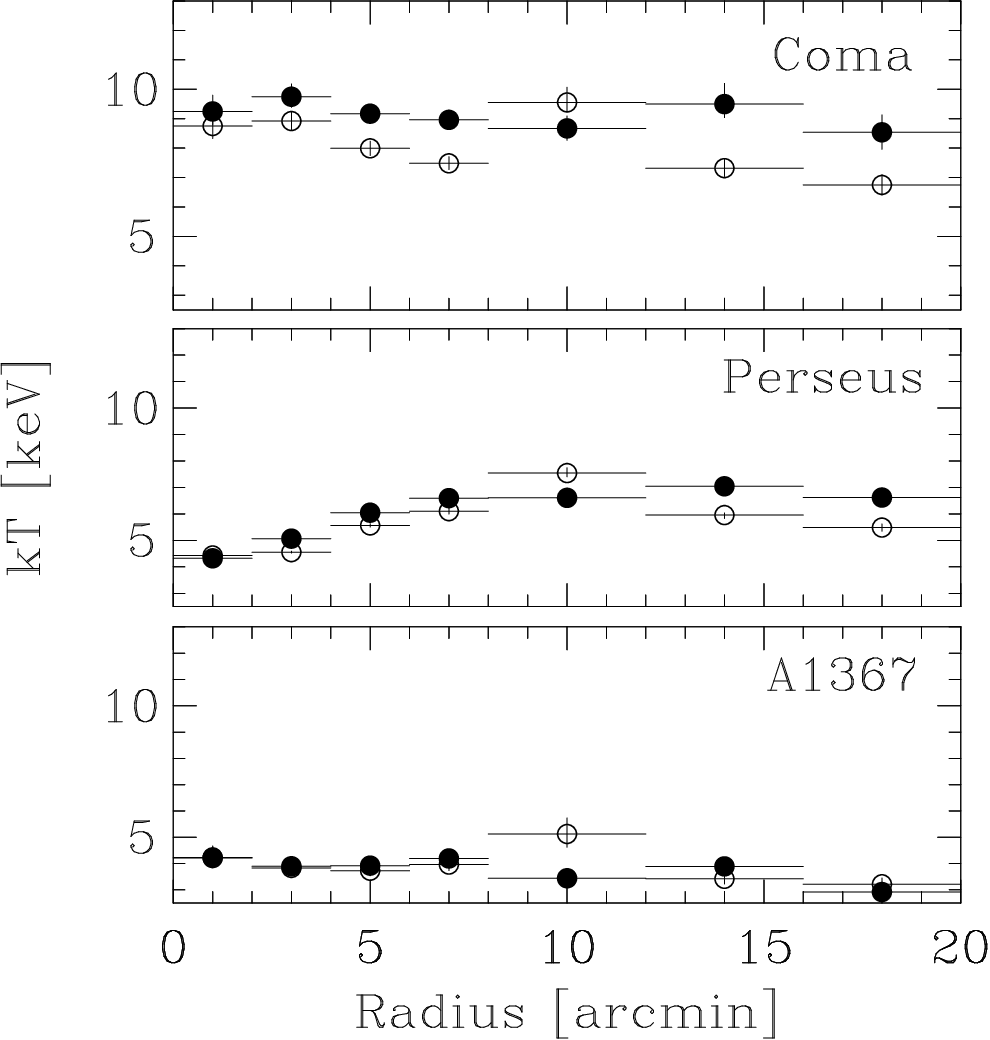,width=4.in,angle=-90}} 
\figcaption{Radial temperature profiles for Coma, Perseus and A1367 
as a function of radius in arcminutes derived by analyzing the
BeppoSAX data using the corrected (filled circles) and uncorrected
(open circles) effective areas. 
}
\vskip 1.cm
\epsscale{0.8}
\centerline{\psfig{file=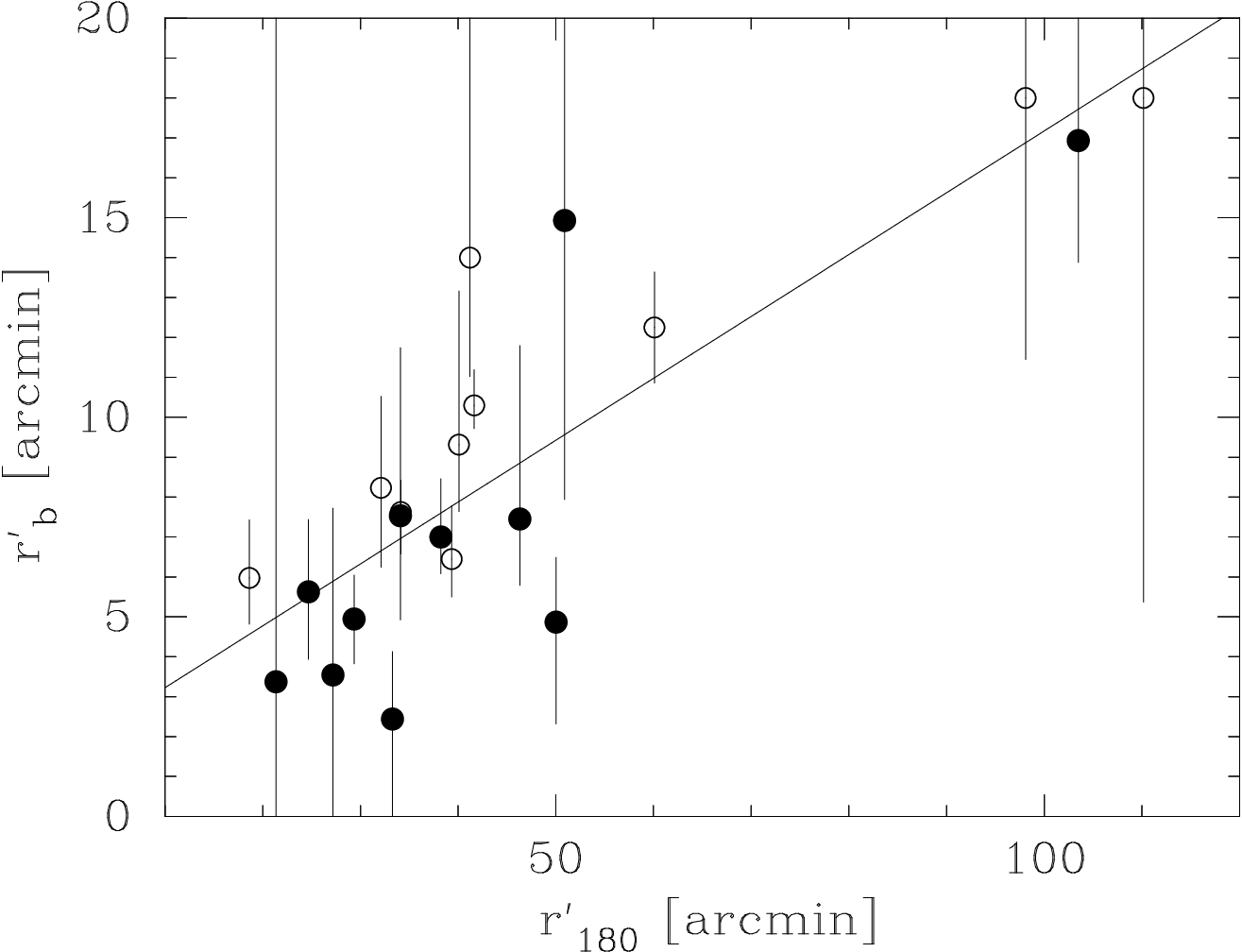,width=4.5in,angle=-90}} 
\figcaption{Break radius in arcminutes, $r^\prime_b$, computed as
described in Section 4.1.6, versus $r^\prime_{180}$ in arcminutes.
Filled circles are CF clusters, open circles are non-CF clusters. }
\end{figure}

\clearpage 

\begin{figure}
\epsscale{0.8}
\centerline{\psfig{file=f9.eps,width=4.5in,angle=-90}} 
\figcaption{Break radius in units of $r_{180}$ as a function of 
redshift for all the objects in our sample. 
Filled circles are CF clusters, open circles are non-CF clusters. 
}
\vskip 1.cm
\epsscale{0.8}
\centerline{\psfig{file=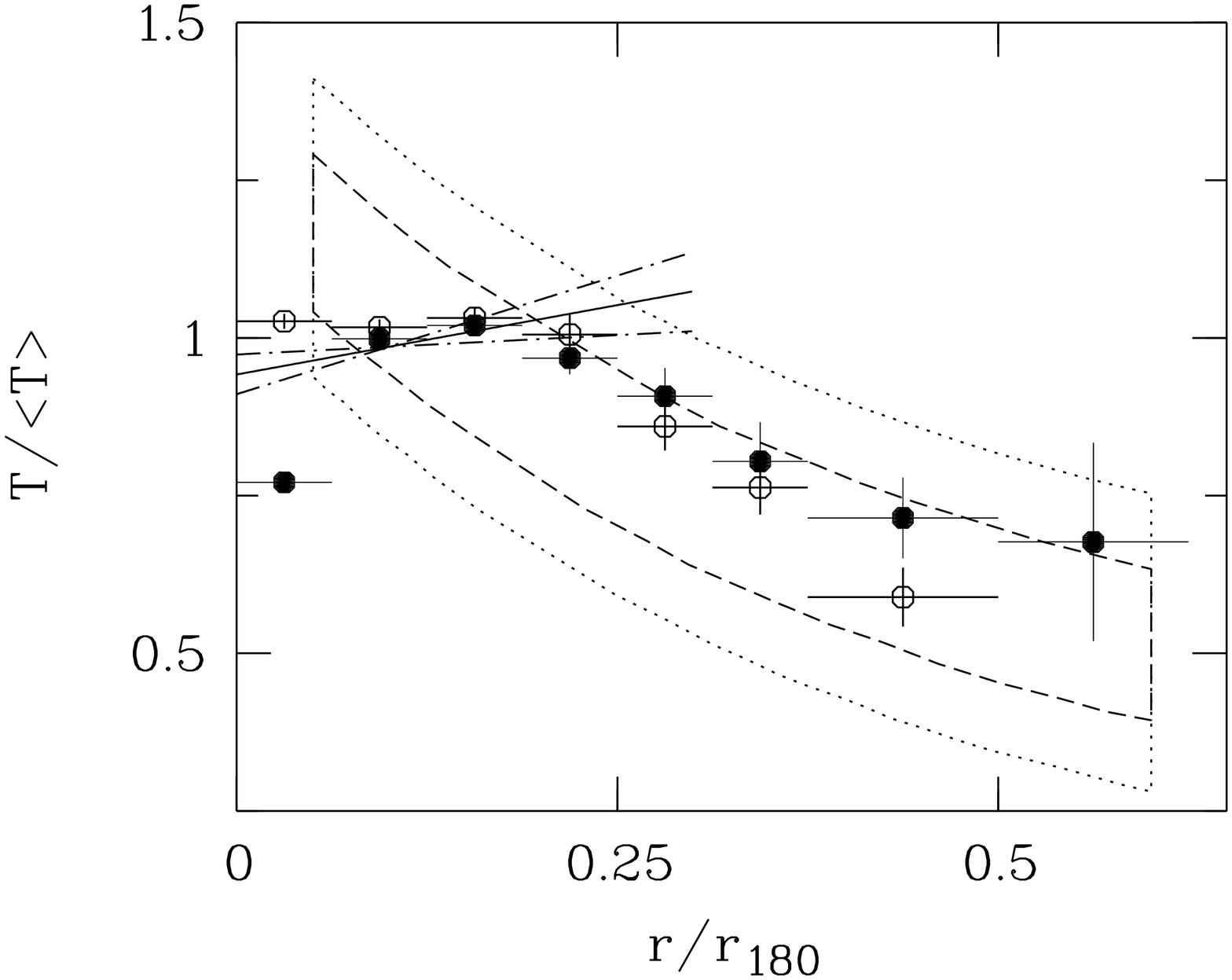,width=6.in,angle=-90}} 
\figcaption{Mean error-weighted temperature profiles for the CF (filled 
circles) and non-CF (open circles) clusters as a function of the
normalized radius. Error bars represent the $1\sigma$ errors. The
solid line is the best-fit linear function computed by Irwin \&
Bregman (2000) with the dot-dashed lines representing the $90\%$
confidence levels of the slope. Also shown is the result in the
composite profile found by Markevitch et al. (1998): the dotted box
encloses approximately all their temperature profiles and most of the
associated error bars, whereas the long-dashed box encloses the
scatter of their best-fit values (see Figure 8 in Markevitch et
al. 1998).}
\end{figure}

\clearpage 

\begin{figure}
\epsscale{0.8}
\centerline{\psfig{file=f11.eps,width=5.in,angle=-90}} 
\figcaption{XMM-Newton temperature profiles (open circles) overplotted to
our BeppoSAX temperature profiles for A1795 (upper panel) and Coma
(lower panel). XMM-Newton data are taken from Arnaud et al. (2001b)
for A1795 and from Arnaud et al. (2001a) for Coma; all $90\%$
uncertainties have been converted into $1\sigma$ confidence limits.}
\vskip 1.cm
\epsscale{0.8}
\centerline{\psfig{file=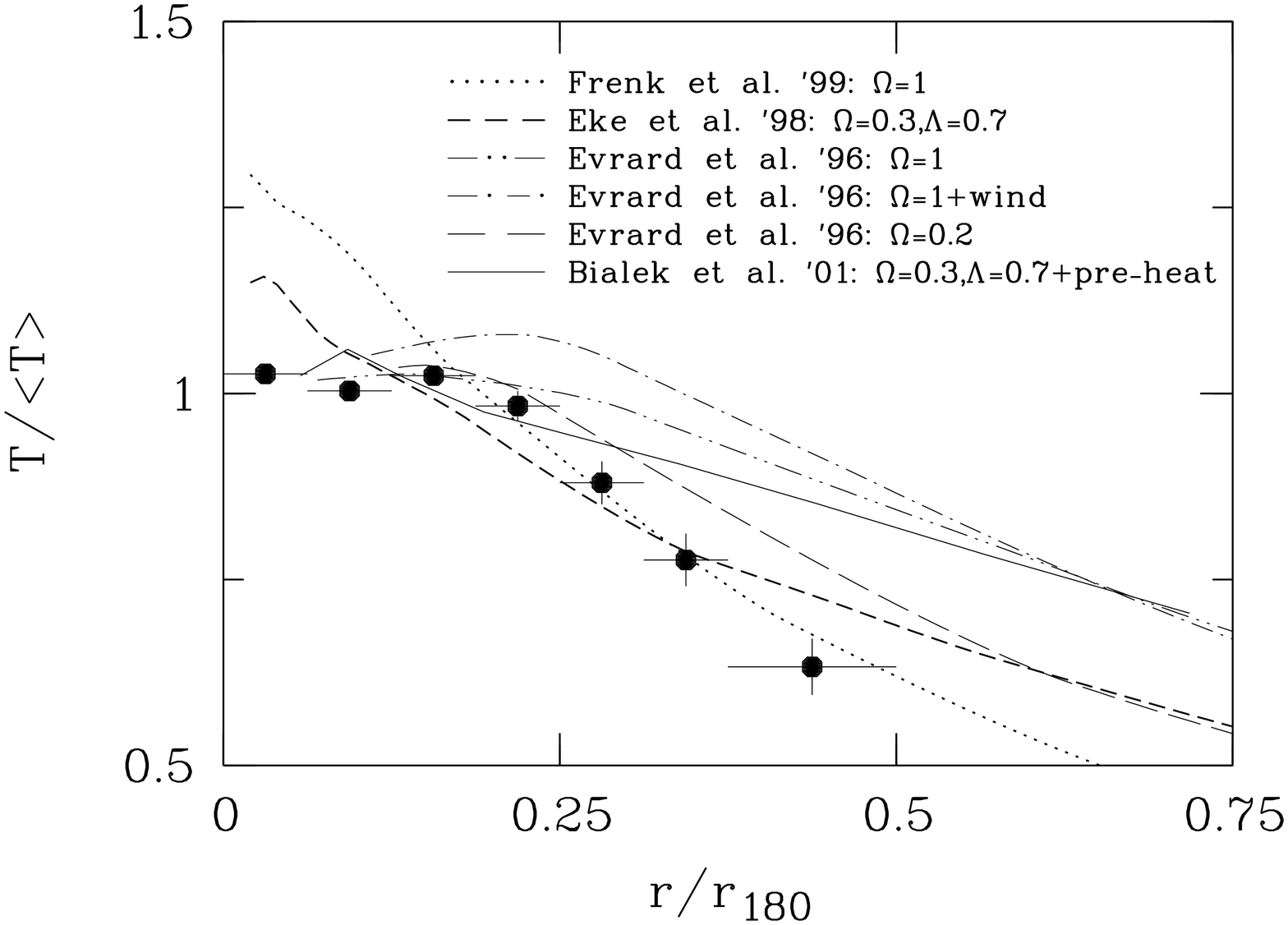,width=6.in,angle=-90}} 
\figcaption{Projected emission-weighted temperature profiles from  
simulations compared to the observed mean temperature profile.  Each
bin of the observed profile (circles), except for the innermost one, has
been computed taking into account all clusters in our sample. In the
case of the innermost bin we have averaged data from non-CF clusters
only. The simulated projected profiles are all computed for CDM
universes and derive from Frenk et al. (1999): $\Omega=1$ (dotted
line); Eke et al. (1998): $\Omega=0.3$ $\Lambda=0.7$ (short-dashed
line); Bialek et al. (2001): $\Omega=0.3$ $\Lambda=0.7$ with initial
entropy-floor S3 (solid line); Evrard et al. (1996): (dot-dot-dashed
line) $\Omega=0.1$, $\Omega=1$+wind (dot-dashed line), and
$\Omega=0.2$ (long-dashed line).}
\end{figure}

\clearpage 

 
\begin{deluxetable}{lrrllr} 
\tabletypesize{\scriptsize} 
\tablewidth{0pt} 
\tablecolumns{6} 
\tablecaption{Observation log for the BeppoSAX cluster sample\tablenotemark{a}} 
\tablehead{ 
\colhead{Target Name} & \colhead{RA(2000)} & \colhead{DEC(2000)} & \colhead{Obs.Date} & 
\colhead{Obs.Code} & Duration \\ 
\colhead{} & \colhead{(degree)} & \colhead{(degree)} & \colhead{yyyy-mm-dd} & 
\colhead{} & \colhead{(ks)} \\
} 
\startdata  
A85            &  10.3750 &  -9.3833 & 1998-07-18 & 60632001 &  93 \\  
A119           &  14.0667 &  -1.2494 & 2000-07-05 & 61091002 & 128 \\  
A426 (Perseus) &  49.9550 &  41.5075 & 1996-09-19 & 60009001 &  80 \\  
A496           &  68.4071 & -13.2619 & 1998-03-05 & 60477001 &  92 \\  
A754           & 137.3421 &  -9.6878 & 2000-05-06 & 60936001 &  62 \\  
               & 137.3375 &  -9.6900 & 2000-05-17 & 61091001 & 123 \\ 
A1367          & 176.1208 &  19.8339 & 1999-12-21 & 60832001 &  97 \\  
A1656 (Coma)   & 194.8950 &  27.9450 & 1997-12-28 & 60126002 &  68 \\  
               & 194.8950 &  27.9450 & 1998-01-19 & 601260021&  24 \\  
A1750          & 202.7188 &  -1.8408 & 2000-01-22 & 60941001 & 101 \\  
A1795          & 207.2080 &  26.5917 & 1997-08-11 & 604080011&  28 \\  
               & 207.2196 &  26.5922 & 2000-01-26 & 60878001 &  93 \\  
A2029          & 227.7313 &   5.7439 & 1998-02-04 & 60226001 &  42 \\ 
A2142          & 239.5833 &  27.2333 & 1997-08-26 & 60169002 & 102 \\ 
A2199          & 247.1592 &  39.5514 & 1997-04-21 & 60169001 & 101 \\  
A2256          & 255.9929 &  78.6419 & 1998-02-11 & 60465001 &  81 \\  
               & 255.9929 &  78.6419 & 1999-02-25 & 60126003 &  51 \\  
A2319          & 290.3025 &  43.9494 & 1997-05-16 & 60226002 &  40 \\  
A3266          &  67.8379 & -61.4444 & 1998-03-24 & 60539002 &  76 \\  
A3376          &  90.4058 & -39.9903 & 1999-10-17 & 60936002 & 110 \\  
A3562          & 203.4100 & -31.6700 & 1999-01-31 & 60638001 &  46 \\  
A3571          & 206.8667 & -32.8656 & 2000-02-04 & 60843002 &  65 \\  
A3627          & 243.5917 & -60.8722 & 1997-03-01 & 60180001 &  34 \\  
2A 0335$+$096  &  54.6458 &   9.9650 & 1998-09-11 & 60675001 & 105 \\  
PKS 0745$-$191 & 116.8792 & -19.2958 & 1998-10-23 & 60539001 &  92 \\   
\enddata 
 
\tablenotetext{a}{Multiple observations of the same cluster have been merged.} 
 
\end{deluxetable} 
 
 
 
\begin{deluxetable}{lccccccccccc} 
\tabletypesize{\scriptsize} 
\tablewidth{0pt} 
\tablecolumns{12} 
\rotate 
\tablecaption{Summary of the BeppoSAX MECS radial temperature profiles.} 
\tablehead{ 
\colhead{Name} & \colhead{kT} & \colhead{kT} & \colhead{kT} &  
\colhead{kT}   & \colhead{kT} & \colhead{kT} & \colhead{kT} & 
\colhead{$<$kT$>$} & \colhead{$\chi^2$/dof} & \colhead{$z$}  & 
\colhead{$r_{180}$} \\ 
\colhead{}                            & \colhead{0$^{\prime}$-2$^{\prime}$}  & 
\colhead{2$^{\prime}$-4$^{\prime}$}   & \colhead{4$^{\prime}$-6$^{\prime}$}  & 
\colhead{6$^{\prime}$-8$^{\prime}$}   & \colhead{8$^{\prime}$-12$^{\prime}$} & 
\colhead{12$^{\prime}$-16$^{\prime}$} & \colhead{16$^{\prime}$-20$^{\prime}$} &\colhead{} & \colhead{} & \colhead{}  & \colhead{Mpc} \\
} 
\startdata  
CF & & & & & & & & & & & \\  
 
A85\tablenotemark{a} & 5.64$\pm$0.13  & 6.87$\pm$0.19 & 6.84$\pm$0.27 
               & 6.98$^{+0.68}_{-0.60}$ & 6.77$^{+1.31}_{-1.04}$ & 
               5.19$^{+1.10}_{-0.83}$ & --- & 6.83$\pm$0.15 & 2.35/4 &  0.0565 &
               3.22 \\ 
 
A426 (Perseus) & 4.33$\pm$0.04 & 5.07$\pm$0.04 & 6.05$\pm$0.08 & 
               6.60$\pm$0.12 & 6.61$\pm$0.17 & 7.04$\pm$0.19 & 6.62$\pm$0.21 & 
               6.68$\pm$0.08 & 4.26/3 & 0.0179 & 3.19 \\ 
 
A496           & 3.56$\pm$0.07 & 4.47$\pm$0.10 & 4.36$^{+0.14}_{-0.13}$ & 4.56$^{+0.23}_{-0.21}$ &  
               4.18$^{+0.44}_{-0.37}$ & 3.76$^{+0.55}_{-0.42}$ & --- & 4.42$\pm$0.08 & 2.55/4 & 0.0329 &  
               2.59 \\  
 
A1795          & 5.59$\pm$0.10 & 6.22$^{+0.16}_{-0.15}$ & 6.27$^{+0.25}_{-0.24}$ &  
               5.75$^{+0.36}_{-0.33}$ & 5.08$^{+0.85}_{-0.73}$ & 3.64$^{+0.80}_{-0.68}$ & --- & 
               6.10$\pm$0.12 & 13.0/4 & 0.0631 & 3.05 \\  
 
A2029          & 7.28$\pm$0.22 & 8.04$\pm$0.35 & 7.46$^{+0.49}_{-0.44}$ & 9.20$^{+1.52}_{-1.13}$ &  
               5.54$^{+1.24}_{-0.93}$ & --- & --- & 7.77$\pm$0.28 & 5.08/3 & 0.0766 &  
               3.44 \\  
 
A2142 & 8.26$^{+0.23}_{-0.22}$ & 8.88$^{+0.30}_{-0.29}$ & 8.79$^{+0.44}_{-0.43}$ & 
                       8.45$^{+0.69}_{-0.66}$ & 6.96$^{+1.03}_{-0.83}$ & 5.96$^{+1.43}_{-0.92}$ &  
                       --- & 8.65$\pm$0.22 & 6.98/4 & 0.0899 & 3.63 \\  
 
A2199 & 4.25$\pm$0.08 & 4.54$\pm$0.09 & 4.59$^{+0.13}_{-0.12}$ & 4.80$^{+0.23}_{-0.19}$ &  
                4.42$^{+0.40}_{-0.38}$ & 4.60$^{+0.65}_{-0.50}$ & 4.32$^{+0.92}_{-0.67}$ & 
                4.62$\pm$0.10 & 1.01/4 & 0.0305 & 2.65 \\ 
 
A3562\tablenotemark{b}   & 5.37$^{+0.42}_{-0.39}$ & 6.00$^{+0.50}_{-0.35}$ & 4.73$^{+0.46}_{-0.37}$ &  
                         4.85$^{+0.80}_{-0.66}$ & 3.47$^{+0.89}_{-0.63}$ & 3.44$^{+0.72}_{-0.59}$ & --- & 
                         4.82$\pm$0.27 & 11.5/4 & 0.0483 & 2.82 \\  
 
A3571          & 7.50$^{+0.27}_{-0.23}$ & 7.55$^{+0.26}_{-0.20}$ & 7.65$^{+0.32}_{-0.26}$ &  
               8.21$^{+0.52}_{-0.50}$ & 6.60$^{+0.62}_{-0.56}$ & 5.96$^{+0.80}_{-0.53}$ &  
               3.87$^{+0.61}_{-0.47}$ & 7.23$\pm$0.17 & 40.5/6 & 0.0391 & 3.32 \\  
 
2A 0335$+$096    & 2.80$\pm$0.04 & 3.39$\pm$0.06 & 3.40$^{+0.10}_{-0.09}$ & 
                 3.67$^{+0.20}_{-0.17}$ &  
                 3.02$^{+0.44}_{-0.47}$ & 2.42$^{+0.38}_{-0.32}$ & --- & 3.38$\pm$0.08 & 9.43/3 & 0.0349 &  
                 2.27 \\ 
 
PKS 0745$-$191   & 7.14$^{+0.15}_{-0.14}$ & 8.52$\pm$0.29 & 7.58$^{+0.53}_{-0.42}$ &  
                 9.07$^{+1.49}_{-1.15}$ & 8.02$^{+3.53}_{-2.10}$ & --- & --- & 8.32$\pm$0.25 & 2.72/3 & 0.1028 &
                 3.56 \\  
 
\hline 
 & & & & & & & & & & & \\  
 
NON-CF & & & & & & & & & & & \\  
 
A119           & 6.55$^{+0.48}_{-0.45}$ & 6.00$^{+0.31}_{-0.25}$ & 6.37$^{+0.34}_{-0.33}$ &  
               5.94$^{+0.40}_{-0.35}$ & 5.73$^{+1.02}_{-0.64}$ & 3.57$^{+0.37}_{-0.34}$ 
               & --- & 5.66$\pm$0.16 & 41.1/5 & 0.0442 & 2.94 \\  
 
A754           & 9.45$^{+0.34}_{-0.30}$ & 9.36$^{+0.28}_{-0.25}$ & 10.40$\pm$0.40 &  
               11.63$^{+0.58}_{-0.54}$ & 11.66$^{+0.99}_{-0.83}$ & 9.08$^{+0.80}_{-0.69}$ &  
               5.04$^{+0.56}_{-0.50}$ &  
               9.42$^{+0.16}_{-0.17}$ & 88.1/6 & 0.0542 & 3.78 \\  
 
A1367          & 4.21$^{+0.43}_{-0.38}$ & 3.90$^{+0.28}_{-0.27}$ & 3.92$^{+0.23}_{-0.22}$ &  
               4.19$^{+0.27}_{-0.25}$ &  3.44$^{+0.35}_{-0.29}$ & 3.89$^{+0.28}_{-0.25}$ &  
               2.92$^{+0.22}_{-0.19}$ & 3.69$\pm$0.10 & 20.3/6 & 0.0220 & 2.37 \\  
 
A1656 (Coma)   & 9.24$^{+0.43}_{-0.38}$ & 10.00$\pm$0.37 & 9.22$\pm$0.24 & 9.23$\pm$0.28 &  
               8.69$^{+0.36}_{-0.35}$ & 9.19$^{+0.41}_{-0.39}$ & 8.47$^{+0.50}_{-0.48}$ 
               & 9.20$\pm$0.13 & 8.67/6 &  0.0222 &
               3.74 \\  
 
A1750          & 4.74$^{+0.41}_{-0.30}$ & 4.25$^{+0.36}_{-0.33}$ & 5.31$^{+0.91}_{-0.73}$ &  
               3.95$^{+0.68}_{-0.51}$ & --- & --- & --- & 4.46$\pm$0.24 & 2.23/3 & 0.0852 &  
               2.52 \\  
 
A2256          & 7.20$^{+0.27}_{-0.26}$ & 7.15$^{+0.20}_{-0.19}$ & 7.23$^{+0.23}_{-0.22}$ &  
               6.93$^{+0.33}_{-0.32}$ & 5.53$^{+0.48}_{-0.43}$ & 4.70$^{+0.60}_{-0.53}$ & --- 
               & 6.97$\pm$0.12 & 25.8/5 & 0.0570 &
               3.26 \\  
 
A2319          & 9.67$^{+0.66}_{-0.48}$ & 9.65$^{+0.55}_{-0.43}$ & 10.15$^{+1.11}_{-0.97}$ &  
               9.76$^{+1.80}_{-1.21}$ & 12.46$^{+2.27}_{-1.76}$ & 10.40$^{+2.21}_{-1.67}$ 
               & --- & 9.82$\pm$$^{+0.37}_{-0.38}$ & 1.66/5 & 0.0560 &   
               3.86 \\  
 
A3266          & 9.22$^{+0.65}_{-0.53}$ & 9.34$^{+0.53}_{-0.42}$ & 9.47$^{+0.67}_{-0.50}$ &  
               8.47$^{+0.72}_{-0.68}$ & 8.30$^{+1.11}_{-0.98}$ & 7.44$^{+1.80}_{-1.38}$ &  
               5.11$^{+1.92}_{-1.23}$ & 8.97$^{+0.29}_{-0.30}$ & 6.82/6 & 0.0550 & 
               3.69 \\  
 
A3376          & 4.23$^{+0.36}_{-0.32}$ & 3.82$^{+0.21}_{-0.18}$ & 4.27$^{+0.27}_{-0.25}$ &  
               3.96$^{+0.33}_{-0.30}$ & 3.48$^{+0.60}_{-0.46}$ & --- & --- & 3.99$\pm$0.13 & 2.92/4 & 0.0455 &  
               2.46 \\  
 
A3627          & 6.29$^{+0.64}_{-0.55}$ & 7.28$^{+0.53}_{-0.44}$ & 6.12$^{+0.36}_{-0.30}$ &  
               7.10$^{+0.50}_{-0.45}$ & 5.59$^{+0.46}_{-0.41}$ & 6.10$^{+0.51}_{-0.38}$ &  
               5.76$^{+0.53}_{-0.44}$ & 6.28$\pm$0.18 & 9.91/6 &  0.0163 &
               3.09 \\  
\enddata 
 
\tablenotetext{a}{Southern subcluster is excluded from  
the spectral analysis.} 
\tablenotetext{b}{A3562 is in the dense Shapley Supercluster and hosts a  
modest cooling flow with a small mass deposition rate of 
$37^{+26}_{-27}$ M$_\odot$ yr$^{-1}$ (Peres et al. 1998).} 
 
\end{deluxetable} 
 
\clearpage 

 
\begin{deluxetable}{llll} 
\tabletypesize{\scriptsize} 
\tablewidth{0pt} 
\tablecolumns{4} 
\tablecaption{Best-fits parameters and $\chi^2$ values for phenomenological models.} 
\tablehead{ 
\colhead{Model}  & \colhead{${T\over <T>} = c$}  
& \colhead{${T\over <T>} = a+b({r\over r_{180}})$}  
& \colhead{${T\over <T>}=  f(t_b,x_b,m)\tablenotemark{a}$}  \\  
\colhead{}  & \colhead{} & \colhead{}  \\ 
\colhead{Sample} & \colhead{$~~~~c$~~~$~~~~~~~~~\chi^2$/dof} & \colhead{~~~~~$a$~~~~~~~~~~~~~$~~~b$~~~~~~~$~~~~\chi^2$/dof} & \colhead{~~~~$t_b$~~~~~~~ $~~~~~x_b$~~~~~~~~ $~~~m$~~~~~~ $~\chi^2$/dof}}
\startdata  
non-CF                  & $1.00\pm 0.01$~~~$231.3/61$  
                        & $1.08\pm 0.01$~~$-0.77\pm 0.08$~~$141.4/60$ 
                        & $1.02\pm 0.01$~~~$0.23\pm 0.01$~~$-2.56^{+0.24}_{-0.25}$~~$87.9/59$ \\  
& & & \\ 
 
CF\tablenotemark{b}     & $1.00\pm 0.01$~~~$82.9/50$ 
                        & $1.07\pm 0.01$~~$-0.55\pm 0.10$~~~~$54.2/49$    
                        & $1.01\pm 0.01$~~~$0.16\pm 0.03$~~$-1.13^{+0.19}_{-0.28}$~~$36.2/48$ \\  
 
& & & \\  
 
all\tablenotemark{b}    & $1.00\pm 0.01$~~~$314.3/112$  
                        & $1.08\pm 0.01$~~$-0.68\pm 0.06$~~$199.8/111$    
                        & $1.02\pm 0.01$~~~$0.20\pm 0.03$~~$-1.79^{+0.29}_{-0.49}$~~$137.4/110$ \\  
\enddata 
\tablenotetext{a}{Broken line model as defined in equation (1).} 
\tablenotetext{b}{Cooling region is excluded from CF clusters on the  
basis of the cooling radius given in Peres et al. (1998).} 
\end{deluxetable} 
 
 
 
\begin{deluxetable}{lllll} 
\tabletypesize{\scriptsize} 
\tablewidth{0pt} 
\tablecolumns{3} 
\tablecaption{Best-fits parameters and $\chi^2$ values for polytropic models.} 
\tablehead{ 
\colhead{Model}  & \colhead{${T\tablenotemark{a}\over <T>} = f(t_o,\gamma,r_c)$} & \colhead{${T\tablenotemark{b}\over <T>} = f(t_o,\gamma)$} \\ 
\colhead{}  & \colhead{} & \colhead{}  \\ 
\colhead{Sample} & \colhead{$~~~~~\gamma^{\tablenotemark{c}}$~~~~~~~$~~~r_c$(Mpc)~~~$~~~\chi^2$/dof}  
                 & \colhead{$~~~~\gamma$~~~~~~~~$~~~~\chi^2$/dof}} 
\startdata  
non-CF                  & $1.67^{+*}_{-0.08}$~~~$1.49^{+0.06}_{-0.13}$~~~$120.7/59$  
                        & $1.09\pm 0.01$~~~~$159.7/60$ \\  
 & & \\ 
CF\tablenotemark{d}     & $1.67^{+*}_{-0.23}$~~~$1.73^{+0.19}_{-0.38}$~~~$45.49/48$ 
                        & $1.06\pm 0.01$~~~~$61.76/49$ \\  
 & & \\ 
all\tablenotemark{d}    & $1.67^{+*}_{-0.07}$~~~$1.54^{+0.08}_{-0.11}$~~~$170.2/110$ 
                        & $1.07\pm 0.01$~~~~$225.9/111$ \\  
\enddata 
\tablenotetext{a}{Polytropic model (as defined in equation 2) with $\beta$ parameter fixed to 2/3 and $\gamma$ constrained between 1 and 5/3.} 
\tablenotetext{b}{Polytropic model (see eq. 2) with $\beta$ parameter fixed to 2/3, core radius fixed to 0.25 Mpc and  
$\gamma$ constrained between 1 and 5/3.} 
\tablenotetext{c}{the $\gamma$ parameter is constrained between the two 
limiting values of 1 (isothermal gas) and 1.67 (adiabatic gas).}
\tablenotetext{d}{Cooling flow region is excluded from CF clusters on the  
basis of the cooling radius given in Peres et al. (1998).} 
\end{deluxetable} 
 

 
\begin{deluxetable}{ll} 
\tabletypesize{\scriptsize} 
\tablewidth{0pt} 
\tablecolumns{2} 
\tablecaption{Best-fits parameters and $\chi^2$ values for polytropic models
applied to radii larger than $0.2~r_{180}$.}
\tablehead{ 
\colhead{Model}  & \colhead{${T\tablenotemark{a}\over <T>} = t_o ({x\over x_o})^{-\mu}$}  \\ 
\colhead{}       & \colhead{}   \\ 
\colhead{Sample} & \colhead{$~~~~~~~\mu$~~~~~~~$~~~T(r=0.3r_c)/<T>$~~$~~~\chi^2$/dof}}  

\startdata  
non-CF & ~$0.92\pm 0.12$~~~~~~~~~~~$0.81\pm 0.02$~~~~~~~~~~$29.57/16$  \\
 &  \\ 
CF     & ~$0.39\pm 0.11$~~~~~~~~~~~$0.84\pm 0.03$~~~~~~~~~~$12.45/23$  \\
 &  \\ 
all    & ~$0.64\pm 0.08$~~~~~~~~~~~$0.82\pm 0.02$~~~~~~~~~~$52.65/39$  \\
\enddata 
\tablenotetext{a}{where $x = r/r_{180}$, $x_o = 0.3$ and $t_o= {T\over <T>}(x=x_o)$. Fits are performed for $x > 0.2$.} 
\end{deluxetable} 
 
\clearpage

\appendix 

\section{Appendix}
 
We assume that at large radii the three-dimensional gas temperature 
and density profiles 
$T(r)$ and $n(r)$ can be described by power-laws of the form: 
$$ T(r) = T_o (r/r_o)^{-\mu} ~~ {\rm and} ~~ n(r) = n_o (r/r_o)^{-\nu}. 
\eqno(A1)$$ 
We approximate the emissivity $\epsilon(r)$ in a given spectral band  
with the expression given in Ettori (2000),  
$$\epsilon(r) =  n^2(r) \lambda T^{\alpha}, \eqno(A2)$$ 
where $\lambda$ is a numerical constant and values for $\alpha$ may be 
found in Table 1 of Ettori (2000). 
The projected emission weighted temperature profile $T(b)$ and surface 
brightness profile $S(b)$, where $b$ is the projected radius, are 
defined as: 
$$T(b) \equiv { \int_0^\infty \epsilon(r) T(r) dl \over  
\int_0^\infty \epsilon(r) dl} ~~~ 
{\rm and} ~~~ S(b) \equiv 2 \int_0^\infty \epsilon(r) dl, \eqno(A3)$$ 
where the integration is along the line of sight, $l$, and the relation 
$r^2 = b^2 + l^2$ is valid.  By substituting equations (A1) and (A2) 
in equation (A3) and by making use of the integration rule: 
$$ \int_0^\infty (l^2 + b^2)^{-z} dl = {\sqrt\pi \over 2}  
{\Gamma(z-1/2) \over \Gamma(z)} b^{1-2z}, \eqno(A4)$$  
where $\Gamma$ is the gamma function, we find that the projected 
temperature and surface brightness profiles can be expressed as 
power-laws of the form: 
$$ T(b) = T^{\prime}_o (b/r_o)^{-\mu^{\prime}} ~~ {\rm and}  
~~ S(r) = S_o (b/r_o)^{-\nu^{\prime}}, \eqno(A5)$$  
where: 
$$ ~~ \mu^{\prime} = \mu ~~~~ {\rm and} ~~~ 
T^{\prime}_o = {\Gamma[1/2(2\nu +(\alpha+1)\mu)-1/2]  
~~\Gamma[1/2(2\nu + \alpha\mu)]\over  
\Gamma[1/2(2\nu +(\alpha+1)\mu)] 
~~\Gamma[1/2(2\nu + \alpha\mu)-1/2]} ~~ T_o,  
\eqno (A6)$$ 
 
$$ ~~ \nu^{\prime} =  2\nu +\alpha\mu - 1 ~~~  
{\rm and} ~~~ S_o = 2\lambda n_o^2 T_o^{\alpha} r_o  
~ {\Gamma[1/2(2\nu +\alpha\mu)-1/2] \over  
   \Gamma[1/2(2\nu +\alpha\mu)]}.  \eqno (A7)$$ 
 
The polytropic index $\gamma$, for a gas described by equation (A1) is
$\gamma = \mu/\nu +1$. Using equations (A6) and (A7) it can be
expressed directly in terms of the observables $\mu^{\prime}$ and
$\nu^{\prime}$:
$$ \gamma ~ = ~ \Big( {\nu^{\prime} \over 2} +   
{1 \over 2} - \alpha\mu^{\prime}\Big)\mu^{\prime}  
+ 1. \eqno(A8)$$  
 
Similar formulae have been derived by Ettori (2000).  In that paper 
the surface brightness profile was described by a $\beta$-model and a 
polytropic relation was assumed between the gas density $n$ and 
temperature $T$. Here we do not assume a polytropic relation to hold 
at all radii (our data shows that this is not the case) and we limit 
ourselves to the outer regions where a power-law behavior holds for 
both the surface brightness profile and the projected temperature 
profile. 
The aim of this Appendix is to show that under the sole assumption 
that the gas density $n(r)$ and temperature $T(r)$ are described by 
power-laws, $n(r)$, $T(r)$ and the polytropic index, $\gamma$, can be derived 
analytically from the parameters obtained by fitting the surface 
brightness profile and the projected temperature profile with 
power-laws. 
 
\end{document}